\documentclass[12pt]{iopart}
\usepackage{iopams}
\usepackage{graphicx}
\usepackage{amsmath}
\usepackage{hyperref}
\usepackage{enumitem}
\usepackage{subcaption}
\usepackage{caption}
\usepackage{pgffor} 
\usepackage{xcolor}
\usepackage{comment}
\def\prd{{\it Phys. Rev.} D~}

\hypersetup{
    colorlinks=true,
    linkcolor=cyan,
    filecolor=magenta,      
    urlcolor=blue,
    citecolor=blue,
}

\usepackage[perpage]{footmisc}

\makeatletter
\renewcommand\@makefnmark{\hbox{\textsuperscript{\normalfont\@thefnmark}}}
\makeatother

\begin{document}

\title[Sequence modeling of higher-order wave modes of binary black hole mergers]{Sequence modeling of higher-order wave modes of quasi-circular, spinning, non-precessing binary black hole mergers}

\author{Victoria Tiki$^{1,2,3}$, Kiet Pham$^{4}$, E.~A. Huerta$^{1,2,3,5,6}$}

\address{$^1$ Data Science and Learning Division, Argonne National Laboratory, Lemont, Illinois 60439, USA}
\address{$^2$ Department of Physics, University of Illinois Urbana-Champaign, Urbana, Illinois 61801, USA}
\address{$^3$ NCSA, University of Illinois Urbana-Champaign, Urbana, Illinois 61801, USA}
\address{$^4$ School of Physics and Astronomy, University of Minnesota, 55455 Minnesota, USA}
\address{$^5$ Department of Computer Science, The University of Chicago, Chicago, Illinois 60637, USA}
\address{$^6$ Department of Astronomy, University of Illinois Urbana-Champaign, Urbana, Illinois 61801, USA}
\ead{vtiki2@illinois.edu}

\begin{abstract}
Higher-order gravitational wave modes from quasi-circular, spinning, non-precessing binary black hole mergers encode key information about these systems’ nonlinear dynamics. We model these waveforms using transformer architectures, targeting the evolution from late inspiral through ringdown. Our data is derived from the \texttt{NRHybSur3dq8} surrogate model, which includes spherical harmonic modes up to $\ell \leq 4$ (excluding $(4,0)$, $(4,\pm1)$ and including $(5,5)$ modes). These waveforms span mass ratios $q \leq 8$, spin components $s^z_{{1,2}} \in [-0.8, 0.8]$, and inclination angles $\theta \in [0, \pi]$. The model processes input data over the time interval $t \in [-5000\textrm{M}, -100\textrm{M})$ and generates predictions for the plus and cross polarizations, $(h_{+}, h_{\times})$, over the interval $t \in [-100\textrm{M}, 130\textrm{M}]$. Utilizing 16 NVIDIA A100 GPUs on the Delta supercomputer, we trained the transformer model in 15 hours on over 14 million samples.  The model's performance was evaluated on a test dataset of 840,000 samples, achieving mean and median overlap scores of 0.996 and 0.997, respectively, relative to the surrogate-based ground truth signals. We further benchmark the model on numerical relativity waveforms from the SXS catalog, finding that it generalizes well to out-of-distribution systems, capable of reproducing the dynamics of systems with mass ratios up to $q=15$ and spin magnitudes up to 0.998, with a median overlap of 0.969 across 521 NR waveforms and up to 0.998 in face-on/off configurations. These results demonstrate that transformer-based models can capture the nonlinear dynamics of binary black hole mergers with high accuracy, even outside the surrogate training domain, enabling fast sequence modeling of higher-order wave modes.
\end{abstract}

\section{Introduction}
\label{sec:intro}

\noindent
Artificial intelligence (AI) and machine learning (ML) are playing an increasingly central role in gravitational-wave (GW) astrophysics, offering powerful tools for tasks ranging from signal detection to population inference. Comprehensive reviews of these developments can be found in~\cite{2025LRR....28....2C, cuoco_review}.
These approaches have enabled the development of low-latency search pipelines~\cite{PhysRevD.111.042010}, robust glitch classification~\cite{BAHAADINI2018172, Chaturvedi:2022suc, glitch_clustering}, and accurate denoising methods~\cite{PhysRevD.108.043024}. Deep learning techniques have matched or exceeded traditional matched-filter performance in detecting binary black hole mergers~\cite{Gabbard:2017lja, Koloniari:2024kww,2018PLBGH,2022PhRvD.105d3006Y,gw_nat_ast,Nousi:2022dwh}, and enabled efficient parameter estimation~\cite{PhysRevLett.127.241103, Green:2020hst} and population-level inference~\cite{PhysRevD.109.064056} orders of magnitude faster than classical methods. 

Sequence modeling of gravitational waveforms has emerged as a promising direction for accelerating waveform generation in gravitational wave astrophysics. Several studies have explored machine learning approaches that map physical parameters to waveforms, particularly focusing on the merger and ringdown regimes. 
For example, Lee et al.~\cite{dl_rg_lee} trained a fully connected network on \texttt{SEOBNRv4} waveforms \cite{PhysRevD.95.044028} to model non-spinning BBH mergers, achieving overlap scores above 99.9$\%$ for the $\ell = m = 2$ mode. This approach accelerated waveform generation and generalized well across parameter space, but was limited to non-spinning systems.
Similarly, Khan and Green\cite{PhysRevD.103.064015} used neural networks to build surrogate models of gravitational waveforms from parameters, achieving mismatches as low as $2\times10^{-5}$ in aligned-spin systems. This work confirmed that simple feed-forward networks could rival interpolation-based surrogates in accuracy with reduced computational cost.
These works exemplify the success of ML-based surrogate modeling but do not directly address the forecasting problem.

\begin{figure}[b]
\centering
\includegraphics[width=0.75\linewidth]{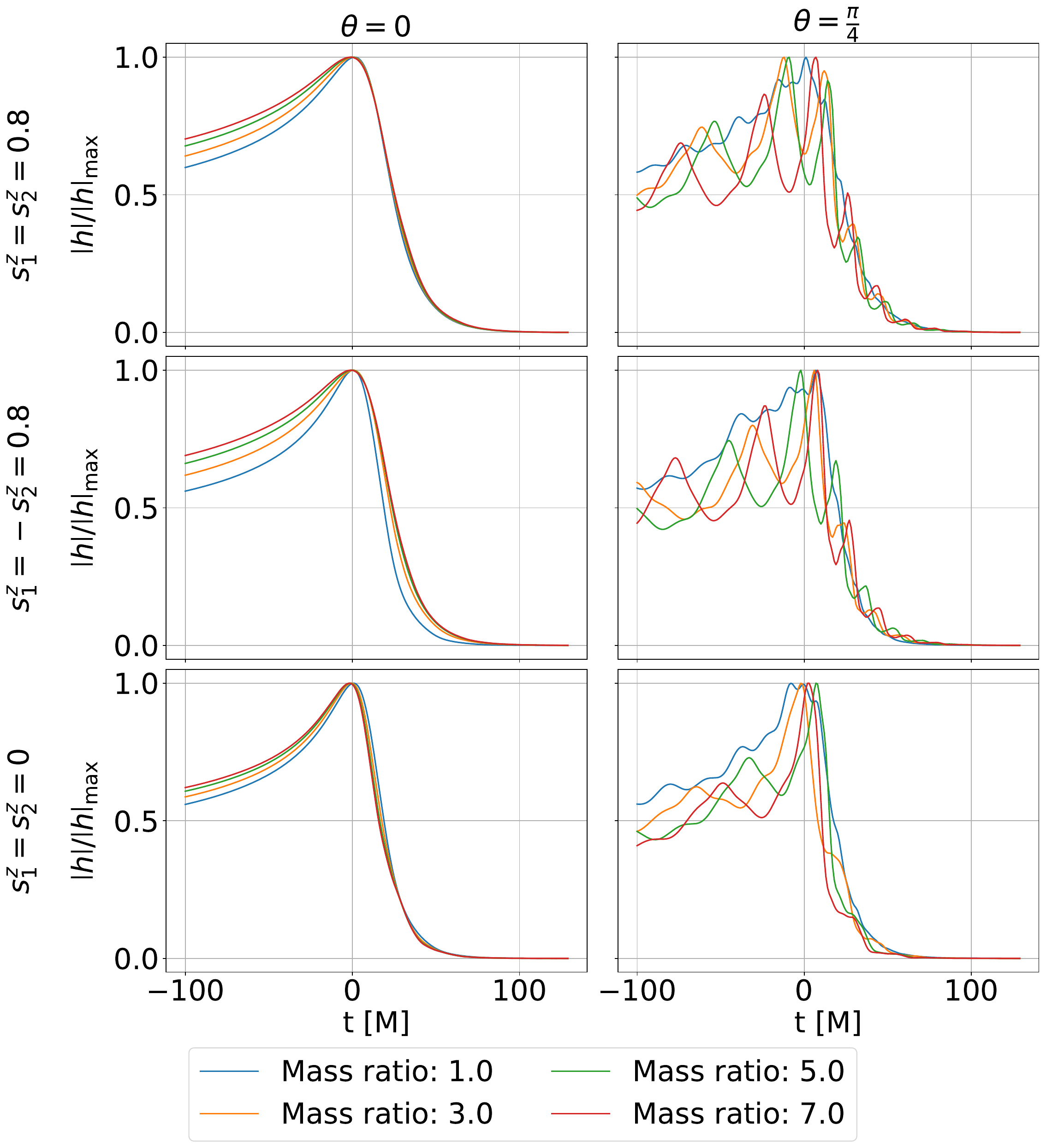} 
\caption{Normalized amplitude $\left| h \right|/\left| h \right|_{\mathrm{max}}$ of gravitational waves for different binary black hole configurations. Notice the impact of higher-order wave modes for inclination angles $\theta = 0$ (left column) and $\theta = \pi/4$ (right column) in terms of amplitude modulations, and the time at which the 
waveform amplitude peaks.}
\label{fig_intro}
\end{figure}

In contrast to parameter-to-waveform models, recent work has explored directly forecasting future gravitational wave signals from earlier parts of the waveform. This task is inherently sequential and lends itself to architectures such as recurrent neural networks or transformer-based architectures. A notable example is~\cite{khan_huerta_forecast}, which introduced a transformer-based model capable of forecasting the late inspiral, merger, and ringdown of signals from spinning, quasi-circular, non-precessing binary black hole mergers. The model was trained on over 1.5 million waveforms generated using the \texttt{NRHybSur3dq8} surrogate\cite{2019PhRvD99f4045V}, and learned to predict late inspiral, merger, and ringdown of the waveform with generally high overlap scores, though with overlaps as low as 0.6 across the test set. The approach was limited to the dominant $\ell=|m|=2$ mode. 
A complementary approach was introduced in~\cite{luo2024multistep}, where a lightweight LSTM network predicted the merger and ringdown using a multi-step sequence strategy. Despite using a much smaller training set based on the \texttt{IMRPhenomD} model \cite{2023ascl.soft07019P}, the LSTM \cite{10.1162/neco.1997.9.8.1735} achieved over 99.6$\%$ mean overlap, demonstrating that compact sequence models can generalize well from limited data—though their scope remained constrained to aligned spins and the dominant harmonic mode. 
More recently, \texttt{BHP2NRMLSur}~\cite{Zhong:2025dwo} introduced a continuous-time neural network that maps perturbation theory waveforms to full NR waveforms across multiple modes and spins. While it achieves high accuracy, it differs from ours in that it transforms approximate templates rather than forecasting waveform evolution from earlier signal segments.

In this work, we extend waveform forecasting to a significantly more complex regime: predicting both polarizations $(h_+, h_\times)$ across higher order harmonic modes, using only early inspiral information from quasi-circular, spinning, non-precessing binary black hole systems. This regime lies deep in the nonlinear sector of general relativity, where spin effects, mode coupling, and gravitational recoil are prominent, and analytical approximations break down. Higher-order modes are particularly important in systems with unequal masses, high spins, or edge-on orientations, where they enhance detection and parameter estimation accuracy~\cite{Abbott:2020tfl,Ng:2023hom}, but also introduce substantial modeling challenges~\cite{Hu:2022modelerror}.
Whereas previous studies~\cite{dl_rg_lee, PhysRevD.103.064015, khan_huerta_forecast, luo2024multistep} focused on the quadrupole mode, we model both GW polarizations $(h_+, h_\times)$ and include all spherical harmonic modes up to $\ell=4$ (including $(\ell,m)=(5,5)$ and excluding $(4,0)$ and $(4,\pm1)$, which are not reliably provided by the surrogate model used for training). Figure~\ref{fig_intro} illustrates the increased complexity introduced by these higher-order modes. The left panel shows representative waveforms from prior studies, where the amplitude evolves smoothly and is dominated by the quadrupole mode. In contrast, our target waveforms (right panel) show non-monotonic structure, strong modulation with inclination angle, and a shifted peak amplitude, demonstrating the richer dynamics that higher-order modes encode.

The key application of this transformer-based model is to serve as a fast surrogate for numerical relativity (NR) simulations, trained on physically realistic waveforms and designed to reflect key domain constraints such as polarization structure and sequence causality.
Rather than functioning as a detection tool, it is designed for systematic modeling of higher-order modes, spin effects, and mode mixing across the BBH parameter space. This enables consistency tests of NR waveforms, waveform interpolation where NR is sparse or unavailable, and rapid hypothesis testing in gravitational theory. 
In doing so, it accelerates the development and validation of analytical waveform models and supports precision studies of strong-field gravity.

To implement this model, we introduce a set of architecture and training design choices that enable accurate and efficient sequence modeling of binary black hole waveforms:

\textbf{1. Data generation and parameter space sampling} We use the \texttt{NRHybSur3dq8} surrogate model to generate gravitational waveforms with higher-order modes up to $\ell = 4$ (including $(5,5)$ but excluding the $(4,0)$, $(4,\pm1)$ modes, which the surrogate does not reliably support). The parameter space covers mass ratios $q \in [1, 8]$, spins $s^z_{{1,2}} \in [-0.8, 0.8]$, and inclination angles $\theta \in [0, \pi]$. To match this domain, we developed HPC-enabled sampling tools to densely cover the space, yielding over 14 million waveforms for training.

\textbf{2. Transformer architecture and distributed training.} To handle the vanishing $h_\times$ in edge-on mergers ($\theta = \pi/2$), we include conditionally activated processing layers that improve accuracy when $h_\times \approx 0$ due to the symmetry of the system. The model forecasts both GW polarizations, $(h_+, h_\times)$, from inspiral through ringdown as distinct channels, using a causal transformer that enforces smoothness and symmetry for physical consistency. Training using data parallelism across 16 NVIDIA A100 GPUs led to convergence within 15 hours on the Delta supercomputer.

\textbf{3. Inference.} Inference on 840,000 test waveforms was completed in under 5 hours using 1 NVIDIA V100 GPU on NCSA's HAL system. The model achieved mean and median overlap scores of 0.996 and 0.997, respectively, when benchmarked against surrogate-model ground truth signals. When evaluated on a small set of numerical relativity waveforms beyond the training domain, the model achieves a mean overlap of 0.969 and a median of 0.902, which increased to a median of 0.998 and a minimum of 0.95 when NR waveforms were constrained to face-on binaries.

\textbf{4. Exploratory interpretability studies.}  To better understand how the transformer achieves accurate predictions, we performed analyses examining the influence of key binary parameters—mass ratio, spin, and inclination angle—on model behavior. These exploratory studies provide insight into which waveform features the model finds predictive and how physical symmetries may be reflected in its learned representations. All code and analysis scripts are available at GitHub~\cite{githubhom} to ensure transparency and reproducibility.

This paper is structured as follows: Section~\ref{sec:methods} describes the data used to train, validate, and test the transformer model, and outlines the model design, training and inference methods. In Section~\ref{sec:res}, we present our findings. Future directions are discussed in Section~\ref{sec:end}. The numerical relativity waveforms employed in the \texttt{NRHybSur3dq8} surrogate model use geometric units, where $G = c = 1$ and $\textrm{M}$ represents the total mass of the binary black-hole merger system. Consequently, all waveform datasets follow this convention, with time measured in units of $\textrm{M}$, where $\textrm{M}$ is the total mass and $1\textrm{M}_\odot = 4.93 \times 10^{-6}$ seconds.

\section{Methods}
\label{sec:methods}

Here we describe the datasets used for this work, the 
approaches used to design and train our transformer model, and 
methods to test its predictive capabilities for sequence modeling.

\subsection{Data}
\label{sec:data}

\noindent Three independent datasets of modeled waveforms were produced for training, validation, and testing of the transformer using the numerical relativity surrogate model \texttt{NRHybSur3dq8}~\cite{2019PhRvD99f4045V}. These waveforms describe the gravitational emission from quasi-circular, spinning, non-precessing binary black hole mergers, encompassing the inspiral, merger, and ringdown phases. The datasets are constrained within the valid parameter space for this NR surrogate model, i.e., mass ratios \(q \leq 8\) and individual spins \(|s_{\{1,2\}}^z| \leq 0.8\). The gravitational wave strain time-series \(h(t, \theta, \phi )\) can be expressed as a sum of spin-weighted spherical harmonic modes, \(h_{lm}\), on the 2-sphere~\cite{newman_penrose}:

\begin{equation}
    h(t, \theta, \phi ) = \sum_{l=2}^{} \sum_{m = -l}^{m = l} h_{lm}(t) {}^{-2}Y_{lm}(\theta, \phi), 
\end{equation}

\noindent where \({}^{-2}Y_{lm}\) are the spin-weight-2 spherical harmonics, \(\theta\) is the inclination angle between the orbital angular momentum of the binary and the line of sight to the detector, and \(\phi\) is the initial binary phase, which we set to zero. We include the higher-order wave modes \(\ell \leq 4\) and \((\ell, m) = (5, 5)\), except for \((4, 0)\), \((4, -1)\) and \((4, 1)\), since \texttt{NRHybSur3dq8} only provides reliable extrapolation for these specific higher-order modes. The waveforms cover the time span \(t = [-5000 \, \textrm{M}, 130 \, \textrm{M}]\).

\begin{figure}[!htbp]
\centering
\includegraphics[width=0.95\linewidth]{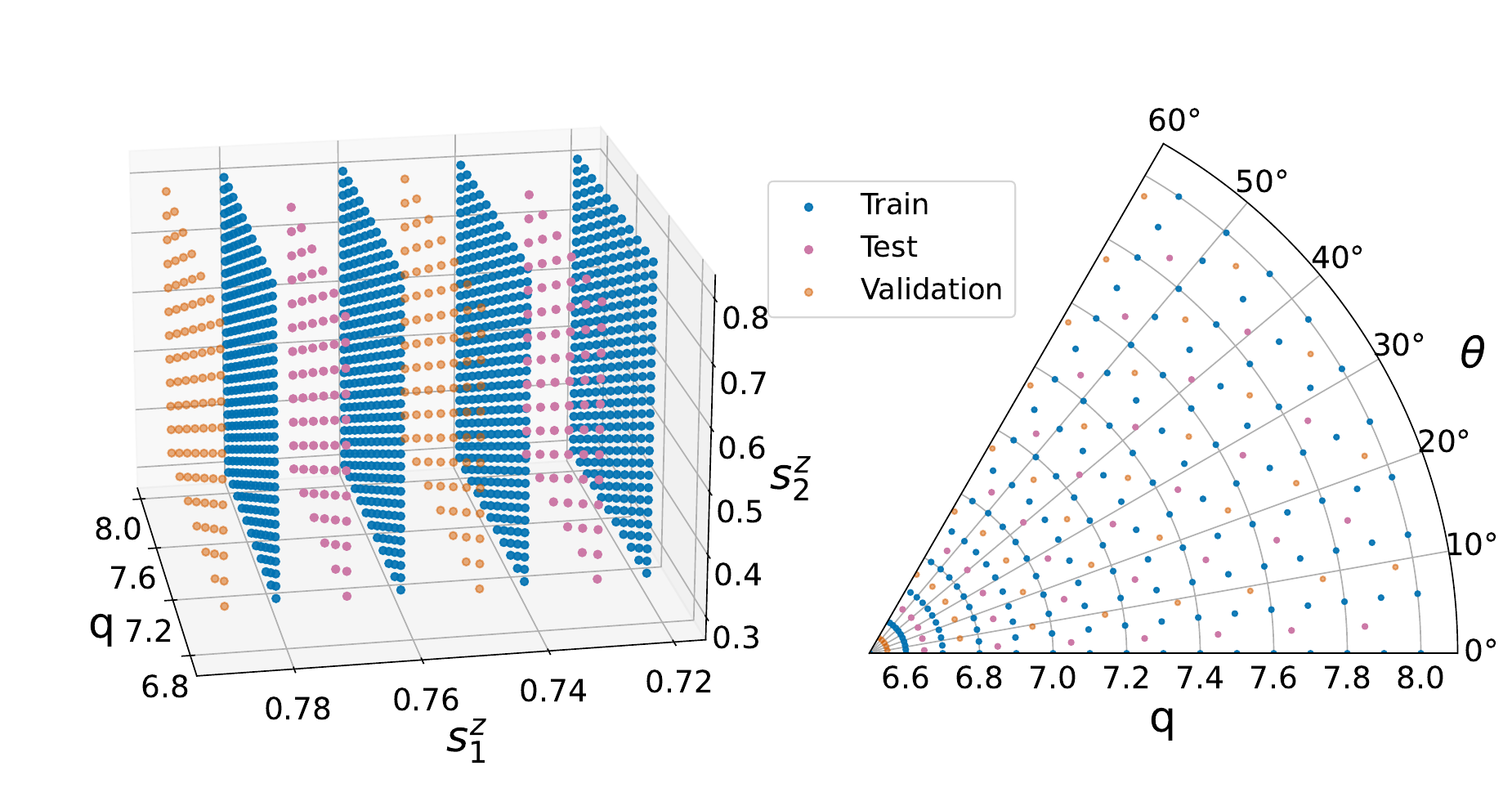}
\caption{Sampling of the signal manifold defined by the mass ratio \(q \in [1, 8]\), aligned spin components \(s^z_{\{1, 2\}}\in [-0.8, 0.8]\), and inclination angle \(\theta \in [0, \pi]\) to generate three independent datasets for training, validation, and testing of our transformer models. The training set uses a uniform grid with step sizes \(\delta q = 0.1\), \(\delta s^z_i = 0.02\), and \(\delta \theta = \pi/29\). The validation and test sets are constructed by interleaving values between training grid points to avoid overlap and evaluate model generalization. For visual clarity, the figure shows a reduced interval of the full signal manifold.}
\label{fig:enter-label}
\end{figure}

\paragraph{Training dataset} The training dataset consists of 14,440,761 waveforms, generated by sampling the mass ratio \(q \in [1, 8]\) in steps of \(\delta q = 0.1\); individual spins \(s_i^z \in [-0.8, 0.8]\) in steps of \(\delta s^z = 0.02\); and inclination angle \(\theta \in [0, \pi]\) in steps of \(\delta \theta = \pi/29\).

\paragraph{Test and validation datasets} Each of these datasets consists of 840,000 waveforms generated by sampling values spaced evenly between the training set values. Figure~\ref{fig:enter-label} illustrates the sampling strategy for generating the training, validation, and test sets, demonstrating that these datasets are independent and do not overlap.

\begin{figure}[!h]
	\centering
	\includegraphics[width=0.7\textwidth,trim={1pt 1pt 1pt 1pt}, clip]{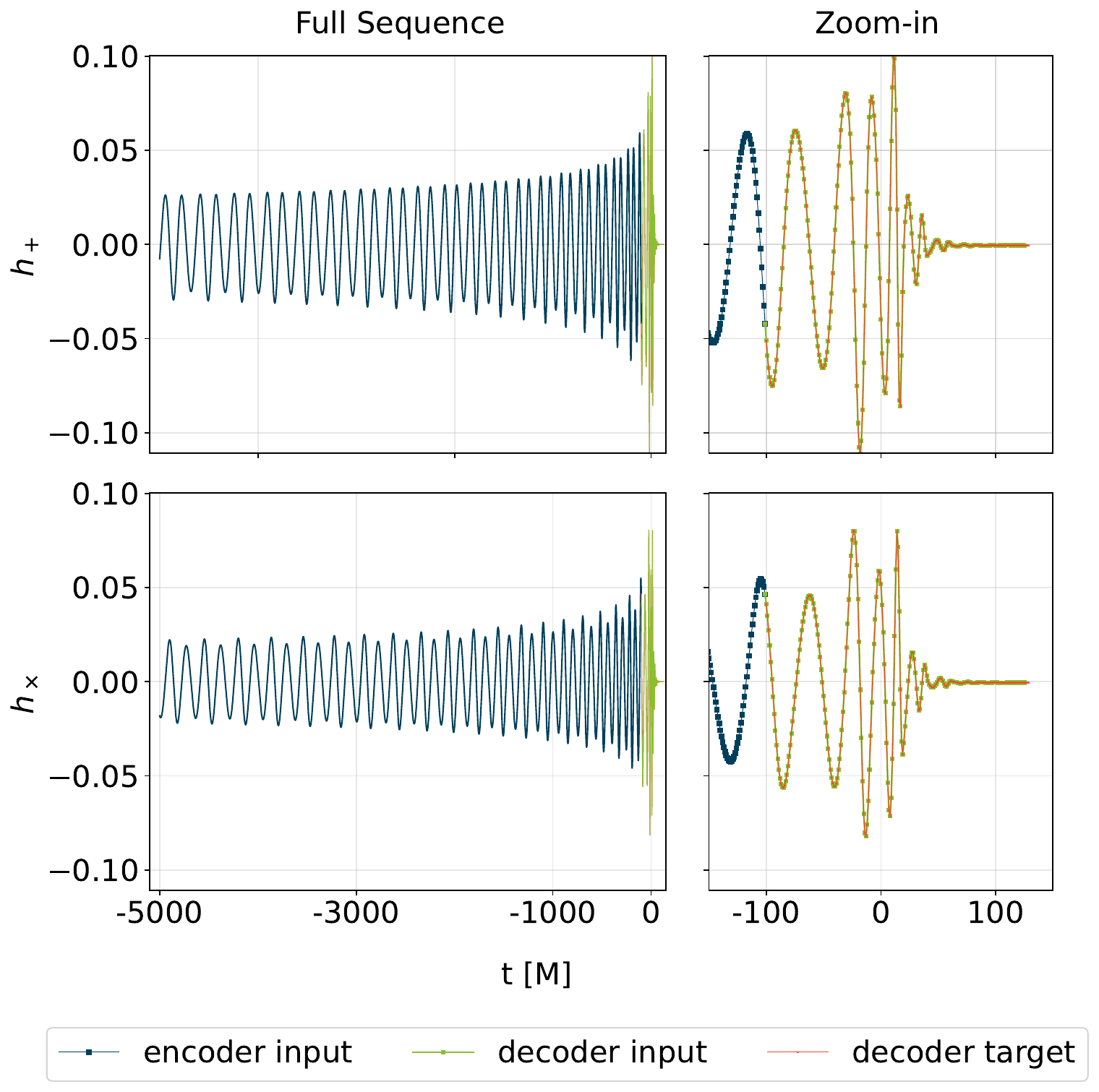}
	\caption{Gravitational wave sequence and separation into encoder input, decoder input, and decoder target for training. Note that the decoder target complements the encoder input exactly, reflecting the transformer's intended function of completing the signal. The decoder input sequence is shifted to the left by one timestep \(\Delta t = 1 \, \textrm{M}\) with respect to the decoder target sequence and thus overlaps the encoder input by one timestep.}
	\label{fig_mom0}
\end{figure}

\paragraph{Sampling and encoder-decoder split} We reduce the temporal resolution of the data by subsampling the original input sequences, selecting every other timestep. This approach reduces computational complexity. The modeled waveform is then split such that the transformer's encoder module receives the early inspiral sequence, consisting of time steps \(t = [-5000 \, \textrm{M}, -100 \, \textrm{M}]\). The decoder module receives the inspiral, merger, and ringdown segments, made up of time steps \(t = [-101 \, \textrm{M}, 129 \, \textrm{M}]\). The target waveform consists of the gravitational wave segment \(t = [-100 \, \textrm{M}, 130 \, \textrm{M}]\), i.e., the decoder target is one timestep ahead of the decoder input, as shown in Figure~\ref{fig_mom0}.

\subsection{Architecture}
\label{sec:architecture}

We extend the original architecture of the transformer model introduced in Ref.~\cite{vaswani2017attention}, which leverages attention mechanisms to weigh different segments of the input sequence and improve performance in sequence prediction tasks. Transformer-based architectures have recently shown strong performance in physical sciences, including stellar light curve analysis\cite{Pan:2024}, gamma-ray burst afterglow modeling\cite{Boersma:2024grb}, and turbulent flow prediction~\cite{Drikakis:2024}. Here, we have modified the model to incorporate properties of gravitational waves. These modifications include an embedding scheme that encodes both polarizations, $(h_+, h_\times)$, as well as an $h_{\times}$ mask that automatically activates additional layers for signals corresponding to edge-on mergers, which are difficult to resolve due to vanishing $h_{\times}$.

\begin{figure}[!htbp]
	\centering
	\includegraphics[width=.9\textwidth, trim={1pt 1pt 1pt 1pt}, clip]{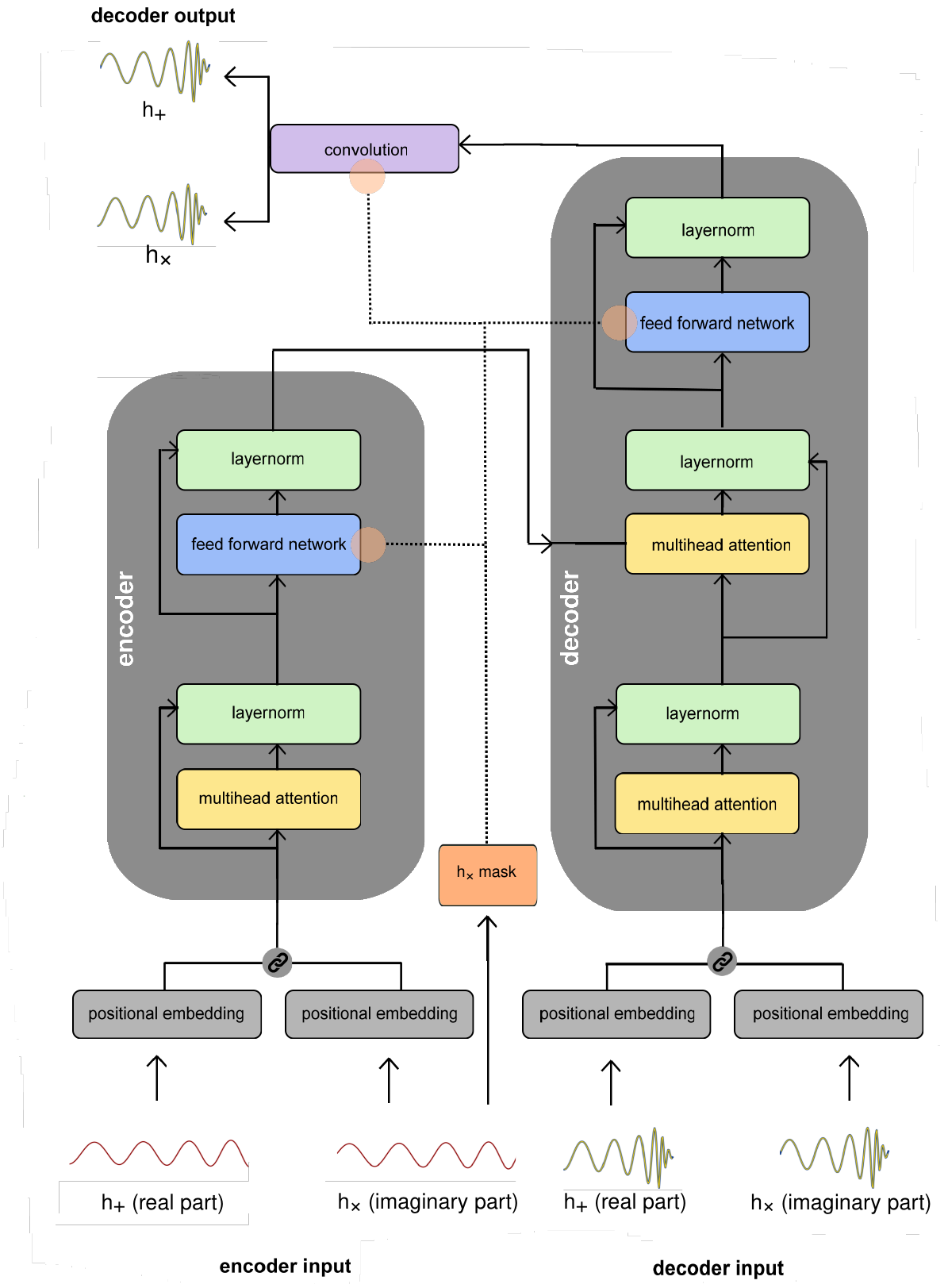}
	\caption{Schematic representation of transformer modules. The input time-series data for \((h_+, h_\times)\) is processed concurrently to generate corresponding predictions \((h_+, h_\times)\). This approach ensures the model maintains physical consistency in forecasting both time-series data.}
	\label{fig:schematic}
\end{figure}

Figure~\ref{fig:schematic} schematically illustrates these modifications. Below, we describe the functionality of each component of the transformer as it processes the gravitational wave sequences. We represent the amplitudes of the input sequences for the encoder and decoder modules by \(x_p\), where \(p\) indicates the position within the sequence of length \(n\). Note that the sequence length \(n\) varies for encoder and decoder inputs, denoted as \(n_{\text{enc}}\) and \(n_{\text{dec}}\), respectively.

\subsubsection{Embedding}
\label{sec:embedding}

The transformer architecture is not intrinsically sensitive to the order of tokens within the input sequence. However, for sequence prediction, information about token order is critical. This issue is addressed by introducing non-trainable positional encoding, which embeds the input sequence into a higher-dimensional space. Through positional encoding, the input sequence \(x_p\) is transformed into a sequence of \(d_{\text{embed}}\)-dimensional vectors, each taking the form:

\[
X_p = (x_p, PE(p, 1), PE(p, 2), \dots, PE(p, d_{\text{embed}-1})),
\]

where \(PE(p, i)\) denotes the positional encoding at position \(p\) and dimension \(i\). This transformation carries information about the original token's position, and we represent this transformed set of vectors as a matrix \(X \in \mathbb{R}^{n \times d_{\text{embed}}}\). The embedding function \(PE\), where \(p\) is the position of any token within the original sequence and \(i\) is the dimension within the positional encoding vector, takes the form:

\begin{align}
PE(p, 2i) &= \sin\left(\frac{p}{10000^{2i/d_{\text{embed}}}}\right), \\
PE(p, 2i+1) &= \cos\left(\frac{p}{10000^{2i/d_{\text{embed}}}}\right).
\end{align}

As demonstrated in Ref.~\cite{vaswani2017attention}, this type of sinusoidal positional encoding enables the model to interpolate and extrapolate positional encodings for unseen positions, providing better generalization than other embedding schemes. Building upon previous transformer models~\cite{khan_huerta_forecast} for gravitational wave forecasting, this work facilitates the processing of complex-valued waveforms: Each input sequence is split into real and imaginary components, corresponding to plus, \(h_+\), and cross, \(h_\times\), polarizations, and independently embedded using the procedure outlined above. These embeddings are then concatenated along the embedding dimension, effectively doubling the input's dimensionality for subsequent processing by the encoder and decoder modules.

\subsubsection{Encoder}
\label{sec:encoder}

The multi-head self-attention mechanism is designed to allow each position in the input sequence to attend to all positions within the same sequence. For each of the \(\mathfrak{h}\, (=10)\) attention heads, the computation reads

\begin{align}
 \text{Attention}(Q^j, K^j, V^j) = \text{softmax}\left(\frac{Q^j (K^j)^T}{\sqrt{d_k}}\right)V^j.
\end{align}

Here, \(Q^j \in \mathbb{R}^{n_{\text{enc}} \times d_q}\), \(K^j \in \mathbb{R}^{n_{\text{enc}} \times d_k}\), and \(V^j \in \mathbb{R}^{n_{\text{enc}} \times d_v}\) denote the query, key, and value matrices, respectively, defined as linear transformations of the input matrix \(X\):

\begin{align}
    Q^j = X W_Q^j \quad K^j = X W_K^j \quad V^j = X W_V^j,
\end{align}

with weight matrices \(W_Q^j \in \mathbb{R}^{d_{\text{embed}} \times d_q}\), \(W_K^j \in \mathbb{R}^{d_{\text{embed}} \times d_k}\), and \(W_V^j \in \mathbb{R}^{d_{\text{embed}} \times d_v}\). In this work, \(d_k = d_q = d_v = d_{\text{embed}} / \mathfrak{h}\).

The matrix \(Q K^T\) functions as a measure of similarity between keys and queries. The softmax activation operates on the rows of \(Q K^T\) and returns weights corresponding to values \(V\). The output of each attention head is concatenated along the last dimension and then linearly transformed:

\begin{equation}
\begin{aligned}
\text{Attention}(Q, K, V) &= \text{concat}\left(\text{Attention}(Q^1, K^1, V^1), \cdots, \right. \\
&\quad \left. \text{Attention}(Q^h, K^h, V^h)\right) W^0,
\end{aligned}
\end{equation}

with \(W^0 \in \mathbb{R}^{d_{\text{embed}} \times d_{\text{embed}}}\), enabling the model to integrate information across different representational subspaces. The resulting attention matrix has the same shape as the input \(X\). After the attention mechanism, each position's output undergoes processing by a position-wise feed-forward network, which applies two linear transformations with ReLU activations. Each sublayer (self-attention and feed-forward network) is enclosed by a residual connection, followed by layer normalization. This configuration aids in stabilizing training.

\subsubsection{Decoder}
\label{sec:decoder}

The decoder receives both the encoder output and the decoder input, which is embedded using the same method as the encoder input. First, a masked multi-head self-attention mechanism is applied to the decoder input only. This mechanism prevents each position in the decoder input from attending to subsequent positions in the sequence, preserving causality and enabling autoregressive predictions during inference. The operation is defined as:

\begin{align}
\text{MaskedAttention}(Q^j, K^j, V^j) = \text{softmax}\left(\frac{Q^j (K^j)^T + M}{\sqrt{d_k}}\right)V^j\,,
\end{align}

where \(M\) is a mask matrix that applies negative infinity to positions not to be attended to, ensuring that the softmax operation assigns them zero weight. As before, \(Q^j\), \(K^j\), and \(V^j\) are the query, key, and value matrices derived from linear transformations on the decoder's input. The dimensionalities of these matrices are the same as those of the encoder self-attention matrices, with the exception of the sequence length \(n\), which varies for encoder inputs and decoder inputs. Outputs from each attention head are again concatenated and linearly transformed. These transformations leave the shape of the decoder input unchanged.

Following the self-attention layer, the cross-attention layer allows the decoder to focus on relevant positions in the encoder output sequence. The queries \(Q\) stem from the previous decoder layer (denoted as \(Y \in \mathbb{R}^{n_{\text{dec}} \times d_{\text{embed}}}\)), while the keys \(K\) and values \(V\) are derived from the encoder output \(X\):

\begin{align}
\text{CrossAttention}(Q^j, K^j, V^j) = \text{softmax}\left(\frac{Q^j (K^j)^T}{\sqrt{d_k}}\right)V^j,
\end{align}

with

\begin{align}
    Q^j = Y W_Q^j \quad K^j = X W_K^j \quad V^j = X W_V^j. 
\end{align}

This operation preserves the shape of the queries, in this case the decoder input, ensuring the sequence length of the decoder output matches that of the decoder input. \(Q K^T\) serves as a measure of similarity between keys and queries. The array returned by the softmax activation function is of shape \(\mathbb{R}^{n_{\text{dec}} \times n_{\text{enc}}}\) and serves as an array of weights for the value array \(V\). Figure~\ref{fig:attention_comparison} presents an instructive visualization of these weights for a choice of input waveform and attention head.

The subsequent processing after the application of the attention mechanisms mirrors that of the encoder module: each position's output is fed to a feed-forward network, residual connections, and layer norms are applied after each attention layer and the feed-forward network to stabilize training. Finally, a 1D convolution is applied to the decoder output to produce predictions for the two time-series \((h_+, h_\times)\) polarizations, respectively.

\subsubsection{$h_\times$ Mask}
\label{sec:mask}

Empirically, we find that separating waveforms with null cross-polarization leads to slightly higher reconstruction accuracy. To facilitate this, we apply a mask to identify and filter waveforms with inclination angle \(\theta = \pi/2\), as these exhibit null cross-polarization. This mask allows us to modify the final convolutional layer and feed-forward layers, which are implemented separately for these waveforms to better capture their distinct waveform structure. The remaining transformations in the model are consistent across all waveforms, ensuring uniform processing while accommodating the distinct features of the null cross-polarization cases. The quantitative impact of this mask is outlined in section \ref{sec:mask_effect}.

\subsection{Training and Inference}
\label{sec:training}

\begin{figure}[!htbp]
\centerline{
	\includegraphics[width=.65\textwidth]{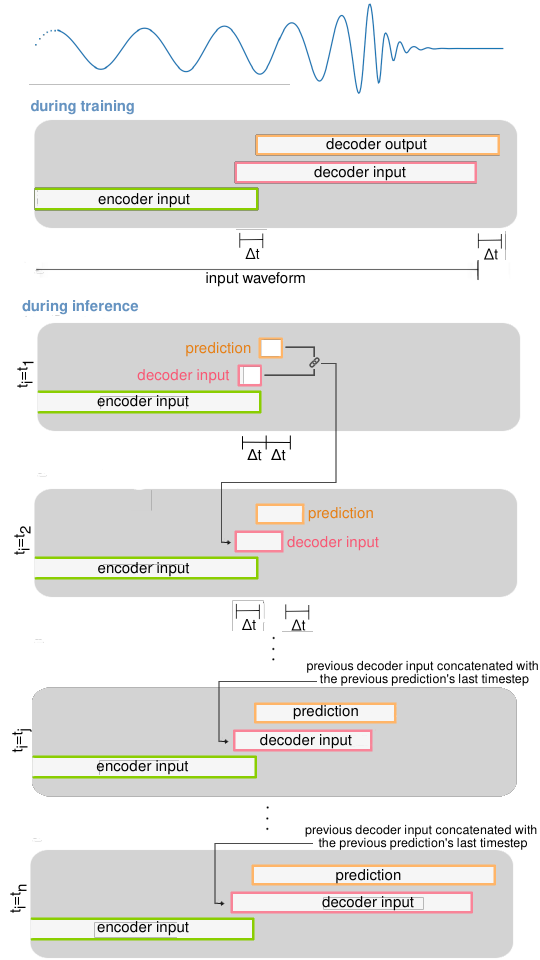}
 }
	\caption{During training, a single decoder output sequence is generated. The attention mechanism uses an attention mask to ensure that each predicted timestep within the decoder output is based on the entire encoder input sequence and the preceding elements in the decoder input. During inference, autoregressive predictions are generated sequentially: at each step, the model produces an output sequence, from which only the last timestep is taken. This output is concatenated with the existing decoder input to form a new input for subsequent predictions. This iterative process continues until the desired sequence is fully generated.} 
	\label{fig_mom2}
\end{figure}

For our transformer model, we used an embedding dimension of 160 (this includes both the contributions from the real and the imaginary part), a feed-forward dimension of 80, and 10 attention heads. The model is trained on a dataset of 14,440,761 waveforms, each consisting of 5,130 total timesteps, with only 362,722 trainable parameters. This results in a high data-to-parameter ratio, which may improve generalization capabilities and help prevent overfitting, as the model is exposed to orders of magnitude more training samples relative to its size.

This model was trained on the NCSA Delta system, using 4 quad NVIDIA A100 GPU nodes. We employed a batch size of 16 and trained the model using the Adam optimizer \cite{kingma2014adam} with a mean squared error (MSE) loss metric. The initial learning rate was set to 0.001, with decay applied during training. Additional dropout or weight decay was not required. Training was distributed across all GPUs using PyTorch's DistributedDataParallel (DDP) framework \cite{2019arXiv191201703P}. The model was trained for 30 epochs and reached convergence in around 15 hours. 

Inference on 840,000 test waveforms was completed in under 5 hours using 1 NVIDIA V100 GPU on NCSA's HAL system. During training, the model processes entire sequences at once. Inference, by contrast, proceeds via sequential autoregressive generation, with each output timestep used to predict the next timestep, which is in turn fed back into the transformer decoder module. The functional differences between training and inference are illustrated in Figure~\ref{fig_mom2}. During inference, the encoder outputs remain constant for each input signal and are therefore computed only once at the beginning of the autoregressive prediction. This approach substantially reduces processing time.

\subsection{Forecasting Efficiency}
To evaluate the inference performance of our transformer-based sequence model for gravitational wave signals, we benchmarked it against the surrogate waveform model \texttt{NRHybSur3dq8}, focusing on waveform generation latency. We selected 50 representative test points within the binary black hole parameter space using a Halton sequence, which ensures uniform coverage and reduces clustering compared to purely random sampling\cite{HALTON1960}. At each point, we generated waveforms 10 times with both models and recorded the per-point average. The surrogate model was run on a single AMD Milan CPU core on the NSCA Delta cluster\footnote{For the surrogate, performance remained consistent for waveform start times as early as -12000M, consistent with prior benchmarks~\cite{PhysRevX.4.031006}.}, while the transformer model was benchmarked on a single NVIDIA A100 GPU on the same cluster. On average, the surrogate generated one waveform in 151 ms (6.6 waveforms/s), and the transformer completed inference for one waveform in 18 ms (55.6 waveforms/s).\footnote{Assuming ideal parallel scaling on one Delta node consisting of 
AMD EPYC 7763 ``Milan'' processors with 64 cores and 4 NVIDIA A100 GPUs, 
these single-instance results suggest a maximum throughput of approximately 422 waveforms per second for the surrogate using all 64 CPU cores, and ~222 waveforms per second for the transformer using all 4 NVIDIA A100 GPUs.}\\
The surrogate model reconstructs waveforms through interpolation of precomputed numerical relativity data, making it computationally lightweight and efficient within its training domain. In contrast, the transformer performs autoregressive sequence forecasting based on learned dynamics of strong-field spacetime evolution, enabling it to generalize beyond the training distribution. While this introduces greater computational complexity per forecast, it also allows the transformer to adapt to previously unseen input conditions. The transformer additionally benefits from GPU-optimized tensor operations and efficient batch processing, whereas the surrogate is more memory-bound and exhibits diminishing returns when parallelized beyond core saturation.

\section{Results}
\label{sec:res}

We evaluated the predictive accuracy of our transformer-based sequence model on a test set comprising 840{,}000 waveforms spanning the full extent of the targeted signal manifold. The accuracy of generated waveforms is quantified using the \emph{overlap} metric, $\mathcal{O}$, which captures the normalized inner product between predicted and true waveforms, maximized over a time shift:

\begin{equation}
\mathcal{O}(h_{\mathrm{true}}, h_{\mathrm{pred}}) = \underset{t_c}{\mathrm{max}} \left( \frac{\langle h_{\mathrm{true}}, h_{\mathrm{pred}}[t_c] \rangle}{\sqrt{ \langle h_{\mathrm{true}}, h_{\mathrm{true}} \rangle \langle h_{\mathrm{pred}}[t_c], h_{\mathrm{pred}}[t_c] \rangle }} \right),
\end{equation}

\noindent where $\langle \cdot, \cdot \rangle$ denotes the inner product over the complex-valued waveform time series, and $h_{\mathrm{pred}}[t_c]$ represents a version of the predicted waveform shifted by $t_c$. The overlap $\mathcal{O} \in [-1,1]$ serves as a surrogate for waveform similarity, with $\mathcal{O} = 1$ indicating perfect agreement.

\subsection{Visual Inspection of Predictions}

\begin{figure}[htbp]
    \centering
    \includegraphics[width=0.7\textwidth]{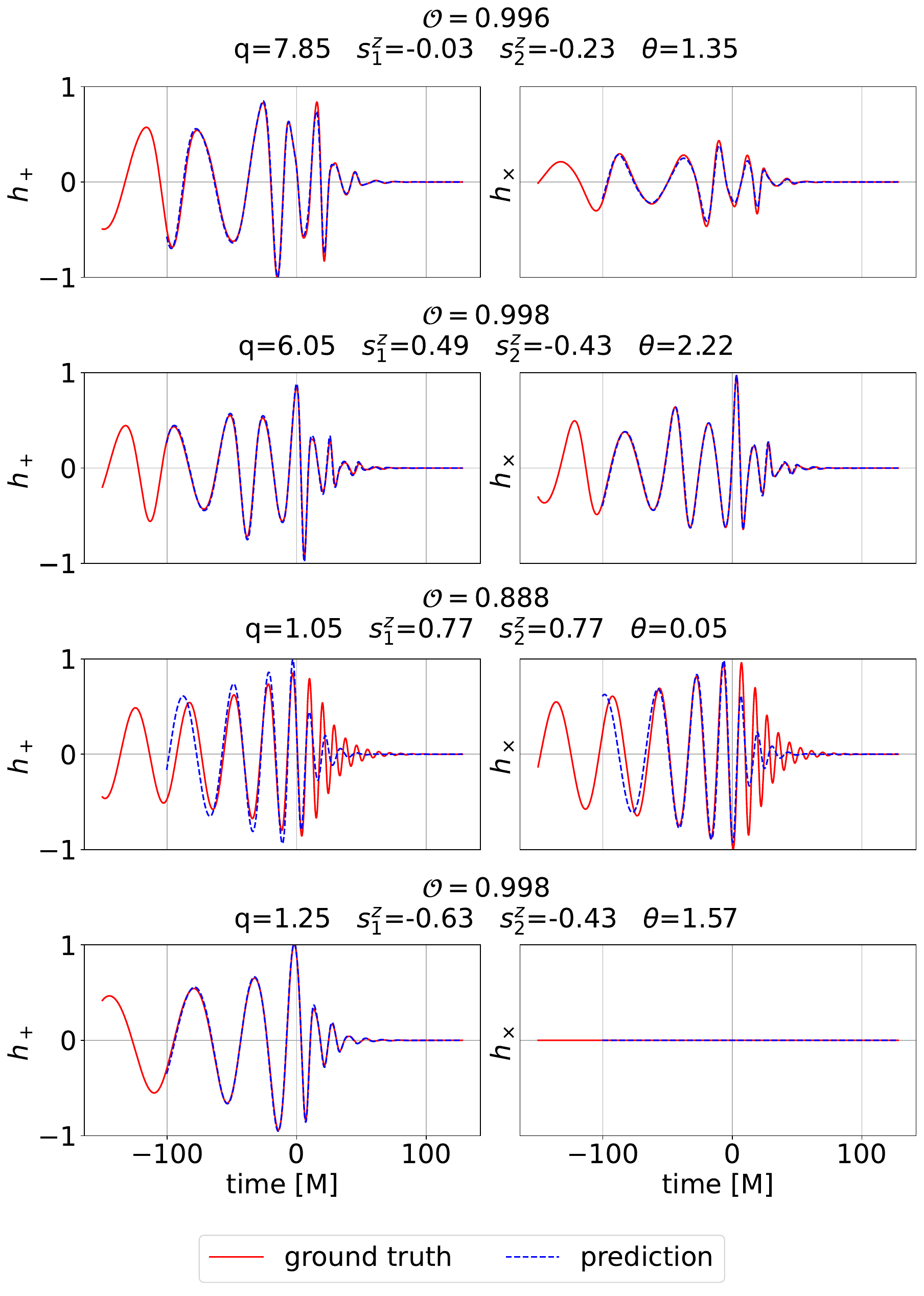}  
    \caption{True and predicted waveforms for selected  \((q, s_1^z, s_2^z, \theta)\) values. The first two rows represent random samples, the third a predicted waveform with one of the lowest overlap values in the test set, and the last a random sample with $\theta=\pi/2$.}
    \label{fig:gallery}
\end{figure}

Figure~\ref{fig:gallery} presents a small sample of model outputs across a range of parameter regimes, compared to the surrogate-based ground truth. The figure includes: (i) two randomly selected waveforms, (ii) one with relatively low overlap, and (iii) one at $\theta = \pi/2$, where $h_{\times} = 0$. These examples provide a visual reference for the range of prediction quality observed across the dataset. For an interactive gallery of more predicted and true waveforms, visit ~\cite{githubhom}. 

\subsection{Overall Performance Across the Signal Manifold}

\begin{figure}[h!]
    \centerline{
    \includegraphics[width=0.85\textwidth]{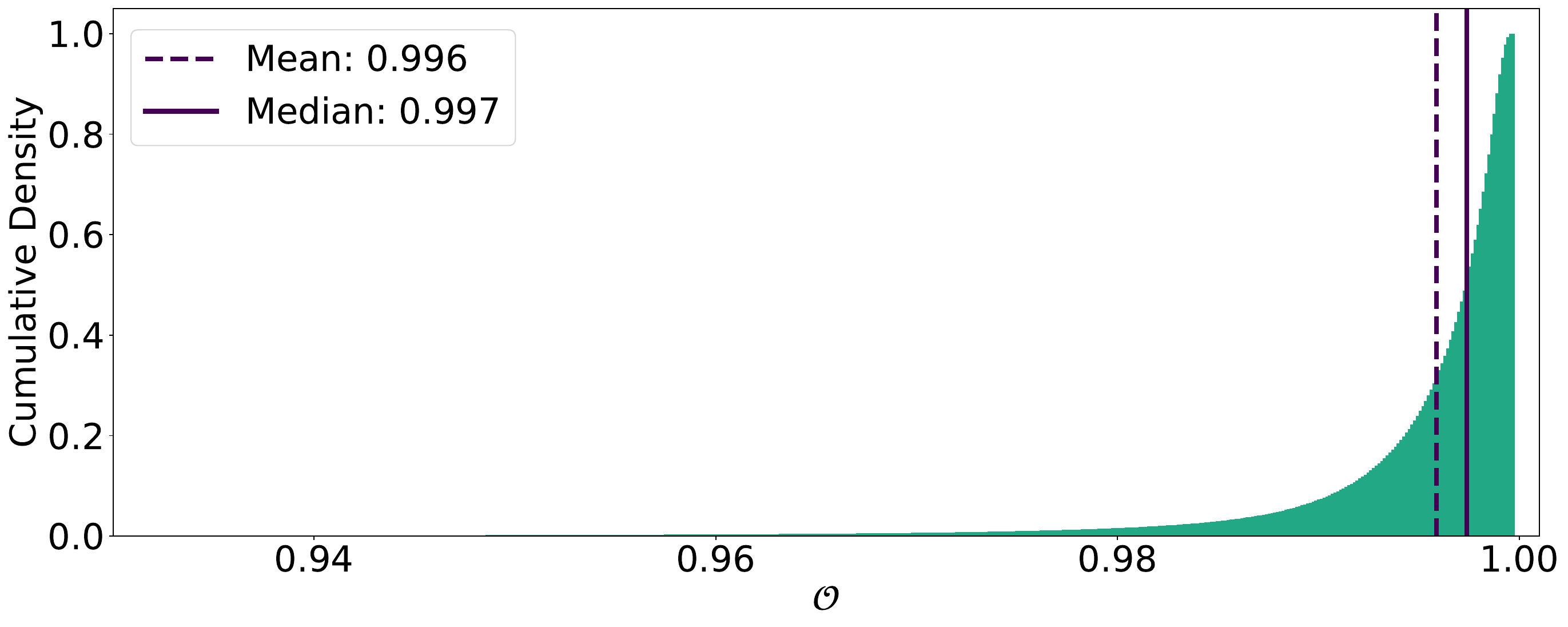}
    } 
    \centerline{
    \includegraphics[width=0.85\textwidth]{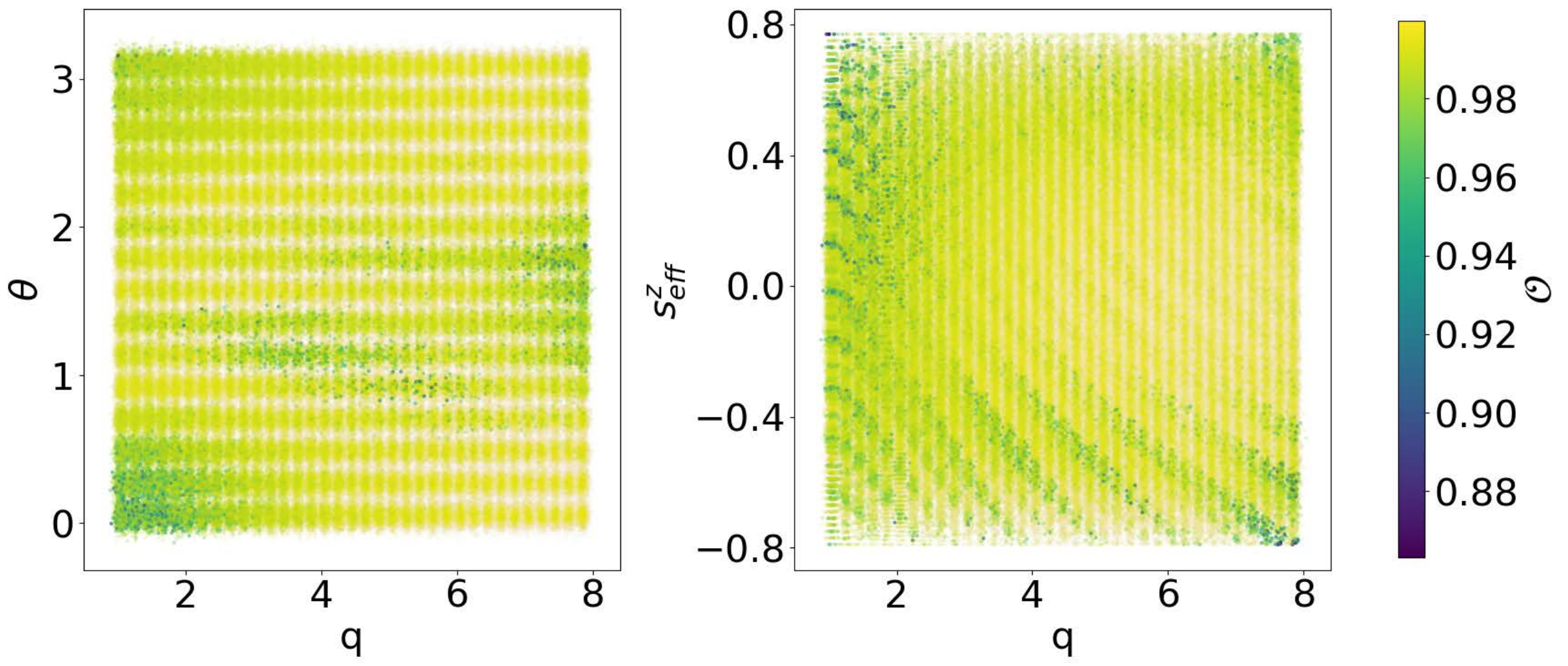}
    }
    \caption{Top panel: Histogram of overlaps. Bottom panels: Overlap $\cal{O}$ as a function of merger parameters. Point opacity is inversely proportional to overlap value, with overlaps close to 1 rendered fully transparent to highlight poor overlaps that might otherwise be obscured. A slight random jitter has been added for the same reason. }
    \label{fig:scatter}
\end{figure}

Figure~\ref{fig:scatter} (top panel) summarizes the distribution of overlap values across the surrogate-based test set. The model demonstrates high predictive fidelity: no overlap value falls below 0.85 across the full test set, fewer than 0.003$\%$ of waveforms yield overlaps below 0.90, and only 7.15$\%$ fall below 0.99, indicating that most predictions align closely with the ground truth. The mean and median overlap values are 0.996 and 0.997, respectively, reflecting high predictive accuracy across the test set.

\subsection{Performance on Numerical Relativity Waveforms}
We conducted an additional benchmark using numerical relativity waveforms from the SXS catalog\cite{SXSPackage_v2025.0.15,SXSCatalogPaper_3,SXSCatalogData_3.0.0}. Specifically, we selected all available 
quasi-circular, non-precessing black hole merger simulations, and reconstructed the full strain signal (excluding memory modes) using all available spherical harmonic modes. While our model was trained only on a subset of modes (see section \ref{sec:data}) and within a limited parameter space ($q \leq 8$ and \(|s_{\{1,2\}}^z| \leq 0.8\)), the NR dataset spans a broader range (including systems with $q\leq15$ and \(|s_{\{1,2\}}^z| \leq 1.0\)). We project each of the simulations using a random inclination angle $\theta \in [0,\pi]$. 
We retained only those waveforms that extend to at least $-5000\textrm{M}$ before merger to match the encoder input requirements, leaving 521 waveforms. 

Despite the mismatch in training domain and harmonic content, the model generalizes reasonably well: we find a median overlap of $\cal{O}$ = 0.969 and a mean overlap of $\cal{O}$ = 0.902 between the transformer predictions and the full NR waveforms. As expected, these scores are lower than the surrogate-based test set, but remain high given the broader parameter coverage. 

We find that low-overlap outliers are broadly distributed across inclination angles, but are absent near face-on or face-off configurations ($\theta \approx 0 \text{ or } \pi$). Restricting to systems within $\pi/4$ of those orientations raises the median overlap to 0.990; further narrowing to $\pi/8$ yields a median of 0.998 and a minimum above 0.95—demonstrating exceptional fidelity in detector-favorable regimes. Notably, at any inclination angle, lower overlaps are not concentrated in out-of-distribution regions. 

For a baseline comparison to our transformer-based forecasting model, we consider another machine learning model that reconstructs gravitational waveforms directly from source parameters. While most machine learning models for gravitational waveforms report performance on surrogate data or synthetic signals, few are evaluated directly on full numerical relativity waveforms. One exception is \texttt{NRSurNN3dq4} \cite{GramaxoFreitas:2024bpk}, a neural network surrogate trained on the $(2,2)$ mode of aligned-spin, non-precessing BBH waveforms with mass ratios up to $q=6$ and dimensionless spin components $\in [-0.99,0.99]$, using the \texttt{NRSur7dq4} surrogate \cite{Varma:2019csw}. It was then fine-tuned and evaluated on a filtered set of 381 SXS waveforms within the same parameter range, achieving a mean overlap of 0.995 and a minimum overlap of 0.979 on a held-out NR test set. Their model differs from ours in mode content, parameter space, and modeling objective (parameter-to-waveform mapping versus time-series forecasting).

Our results suggest that the model retains considerable generalization capacity even when applied to physical systems and mode structures beyond those it was explicitly trained on.

\subsection{Dependence on Physical Parameters}

To examine how predictive accuracy on the surrogate-based test set varies across the parameter space, we first visualize overlap scores as a function of key physical parameters. The bottom panels of Figure~\ref{fig:scatter} display two scatter plots: one mapping overlap against mass ratio and inclination angle $(q,\theta)$, and the other against mass ratio and effective spin $(q,s_{\mathrm{eff}}^z)$. In both plots, point opacity is inversely scaled with overlap value, making regions of lower reconstruction accuracy visually prominent. These visualizations reveal several trends:

\begin{itemize}
    \item \textbf{Mass ratio ($q$):} Degradations are observed at the edges of 
    the parameter space for $q\sim\{1,8\}$.
    \item \textbf{Inclination angle ($\theta$):} Reduced 
    performance is localized near $\theta \sim \{0,\pi/2\}$, corresponding to face-on and edge-on configurations, respectively.
    \item \textbf{Effective Spin ($s_{\mathrm{eff}}^z$):} Defined as
    \begin{equation}
    s_{\mathrm{eff}}^z = \frac{q s_1^z + s_2^z}{1 + q},
    \end{equation}
    a standard parameter used in gravitational-wave population modeling \cite{PhysRevLett.116.241102}. Highly anti-aligned spins (i.e., lower values of $s_1^z$, the spin component of the primary black hole) are associated with a reduction in reconstruction quality, with these regions visible as distinctive green bands in the bottom right panel of Figure~\ref{fig:scatter}.
\end{itemize}
We provide additional results in~\ref{appendix:slices}, where we study the behavior of the transformer by partitioning the signal manifold into fixed spin slices across the \((q, \theta)\) plane.

\subsection{Importance of the $h_{\times}$ Masking Mechanism}
\label{sec:mask_effect}

An important component of this model is the application of a conditional masking mechanism that affects edge-on orientations where the cross polarization $h_{\times} \approx 0$. Ablation tests excluding this mechanism resulted in a decrease in average overlap to 0.994, and a drop in minimum overlap to 0.585 on the surrogate-based test set. These results underscore the importance of the $h_{\times}$ mask for maintaining robustness across observational scenarios. Excluding this conditional masking mechanism yields a $\sim15\%$ inference speedup, offering a trade-off that can be tuned based on application requirements.

\subsection{Exploratory Interpretability Studies} 

These analyses aim to identify which waveform features the model relies on most during prediction. While interpretability results may reflect aspects of the input encoding or architecture—such as symmetry, vanishing components, or decoder proximity—many of the observed patterns align with known physical principles in gravitational wave modeling. The following studies should thus be viewed as indicative of learned model behavior, with findings potentially offering insight into physically meaningful signal properties.

\subsubsection{Obfuscation Experiments}
\label{obfuscation_experiments}

To understand which regions of the waveform most contribute to predictive performance, we conduct a systematic input obfuscation study. In each test, we zero out a fixed segment of the encoder input—effectively masking that region—and measure the resulting degradation in predictive performance. This allows us to estimate the relative importance of each input interval. For a gallery of predicted waveforms resulting from these obfuscation experiments, see \cite{githubhom}.

\begin{figure}[!htbp]
    \centering
    \includegraphics[width=0.8\textwidth]{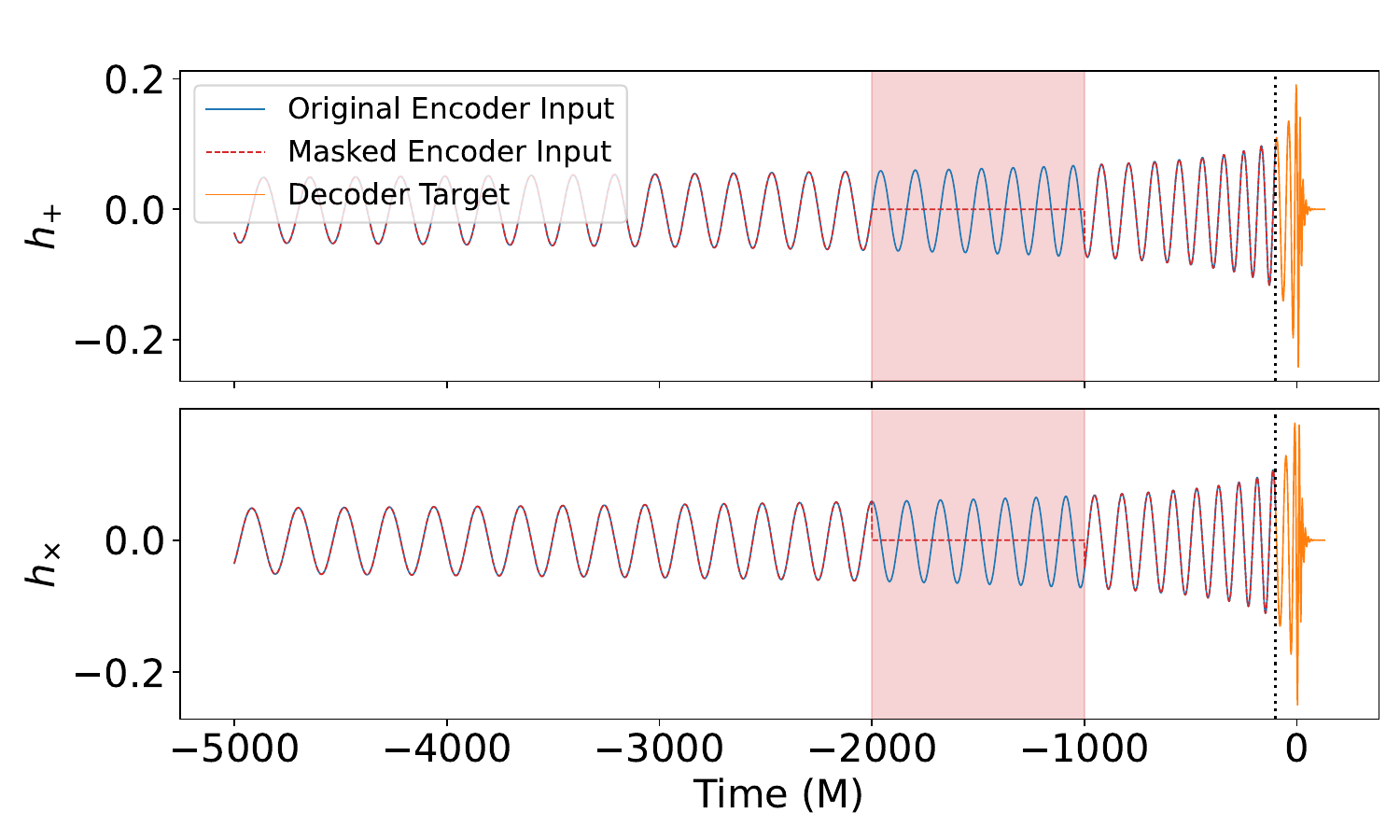}
    \caption{Example of obfuscation: a segment of the input waveform between $t\in[-2000\mathrm{M}, -1000\mathrm{M}]$   is zeroed out before inference. }
    \label{fig:mask_example}
\end{figure}

Figure~\ref{fig:mask_example} illustrates the masking procedure: a segment of the encoder input (blue) from $t\in[-2000\mathrm{M}, -1000\mathrm{M}]$ is replaced with zeros (red), while the decoder target (orange) remains unaltered. This setup ensures that any change in output quality can be directly attributed to the missing input region.

\noindent We tested the following obfuscated intervals: $(-5000\mathrm{M}, -4000\mathrm{M})$, $(-4000\mathrm{M}, -3000\mathrm{M})$, $(-3000\mathrm{M}, -2000\mathrm{M})$, $(-2000\mathrm{M}, -1000\mathrm{M})$, $(-1000\mathrm{M}, -200\mathrm{M})$, and $(-200\mathrm{M}, -120\mathrm{M})$. All segments are approximately 1000$\mathrm{M}$ long to allow fair comparison, except the final one, $(-200\mathrm{M}, -120\mathrm{M})$, where we retain the last 20 timesteps to preserve continuity at the encoder-decoder boundary. Though shorter, this segment lies near the prediction start and, as later analysis shows, has an outsized effect on performance.  This design helps ensure that differences in outcome reflect learned importance rather than artifacts of segment length. 

\begin{figure}[!htbp]
    \centering
    \includegraphics[width=0.9\linewidth]{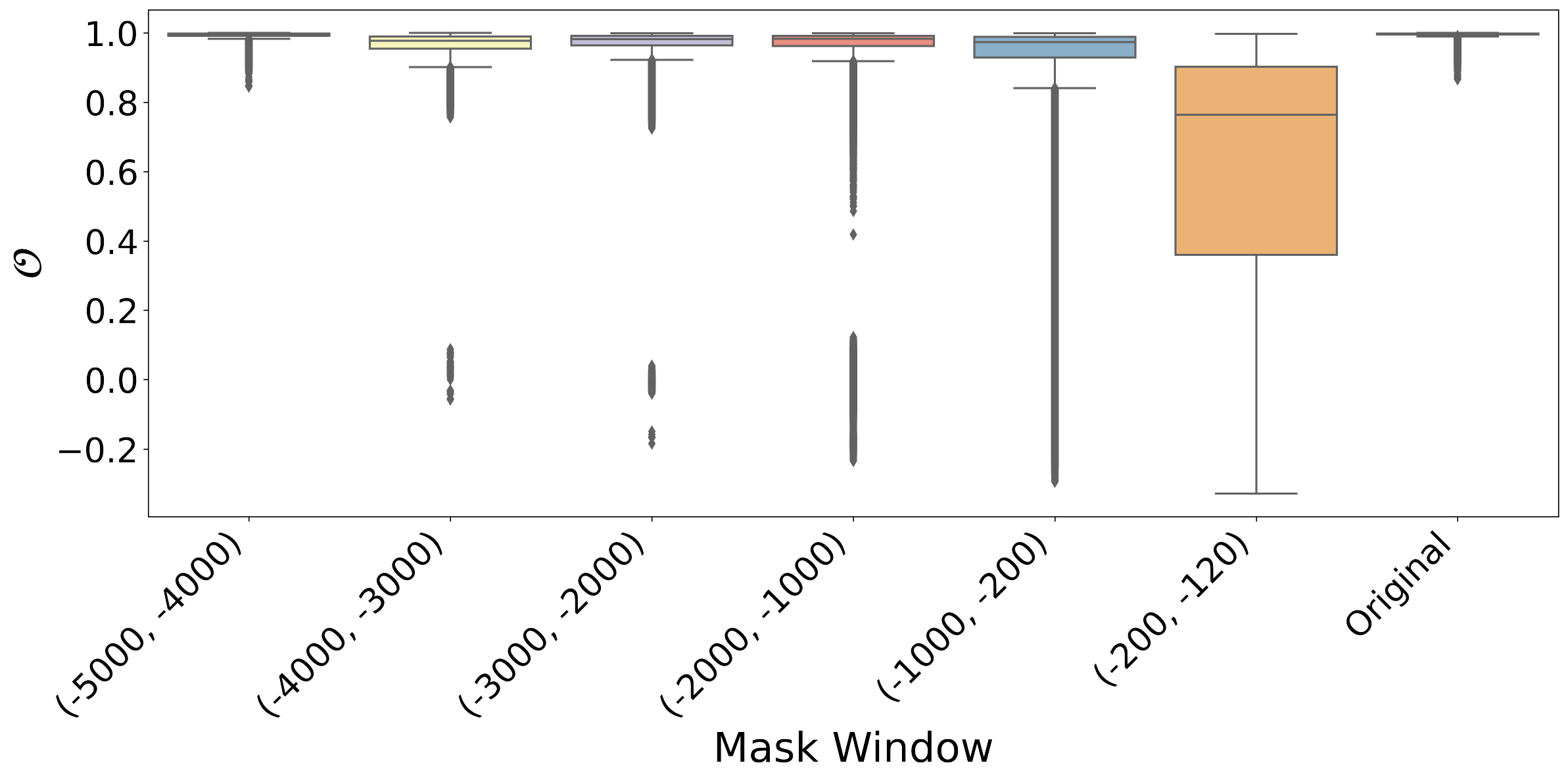}
    \caption{Distribution of overlap scores $\mathcal{O}$ for different obfuscated input intervals. Each box shows the interquartile range, with the median marked by a central line. Whiskers extend to $1.5 \times$ IQR, and outliers are shown as individual points.}
    \label{fig:boxplot}
\end{figure}

To quantify performance degradation, we compute the overlap score $\mathcal{O}$ between predicted and ground truth waveforms. Figure~\ref{fig:boxplot} summarizes the distribution of $\mathcal{O}$ values across all obfuscation intervals. Unsurprisingly, early inspiral segments have an overall limited impact on prediction quality, whereas regions approaching merger are disproportionately influential. Masking the final 200$\mathrm{M}$ results in a decline in overlap and an increase in variance, underscoring the role of late-time dynamics in informing accurate reconstructions.

To probe how this sensitivity varies with physical parameters, we calculate the Pearson correlation coefficient between the overlap drop (i.e., performance loss from masking) and each system's intrinsic properties: mass ratio $q$, spin components $(s_1^z, s_2^z)$, and inclination angle $\theta$. Since inclination effects are approximately symmetric around $\theta=\pi/2$, we use the transformed variable $\lvert \theta- \pi/2 \rvert$ to better reflect this structure. The resulting correlation matrix is shown in Figure~\ref{fig:correlation_matrix}. While correlation values provide a coarse summary of parameter importance, they can obscure nonlinear relationships. \ref{appendix:drop_bins} shows binned overlap drops, revealing structure not captured by Pearson correlation alone.

The correlations reveal several noteworthy trends:

\begin{itemize}
    
    \item \textbf{Mass Ratio ($q$):} A moderate negative correlation appears in the earliest segments, suggesting that equal-mass binaries carry more predictive utility in the early inspiral compared to asymmetric systems. While high mass-ratio systems exhibit richer harmonic content due to enhanced subdominant modes, equal-mass systems produce smoother, more symmetric signals dominated by the quadrupole mode \cite{Berti:2007fi}. These smoother signals may be more readily captured and generalized by the network in this regime. The features of asymmetric binaries become more prominent, and more predictive, closer to merger, as reflected in the steadily increasing match drop with $q$ in later segments.

    \item \textbf{Spin of the More Massive Black Hole ($s_1^z$):} The correlation with overlap drop is strongest in the final input segment ($-200\textrm{M},-120\textrm{M}$), suggesting that the model relies more heavily on late-time features when $s_1^z$ is aligned with the orbital angular momentum. This is in line with previous works which show that spin effects are most prominent near merger, where relativistic velocities enhance spin–orbit and spin–spin couplings. These contribute at higher post-Newtonian orders and thus have limited influence during early inspiral. Near merger, high spins affect phase evolution and waveform amplitude, as reflected in the amplitude's dependence on final spin and the emergence of spin-driven multipolar radiation \cite{Burke:2019yek,refId0}. 

    \item \textbf{Spin of the Less Massive Black Hole ($s_2^z$):} Correlations with $s_2^z$ are weak across all intervals, indicating a diffuse and relatively minor influence on waveform morphology. This is consistent with post-Newtonian predictions that attribute only a minor role in phase evolution to the spin of the smaller object~\cite{Blanchet:2013haa}. 
    
\begin{figure}[h]
    \centering
    \includegraphics[width=0.95\textwidth]{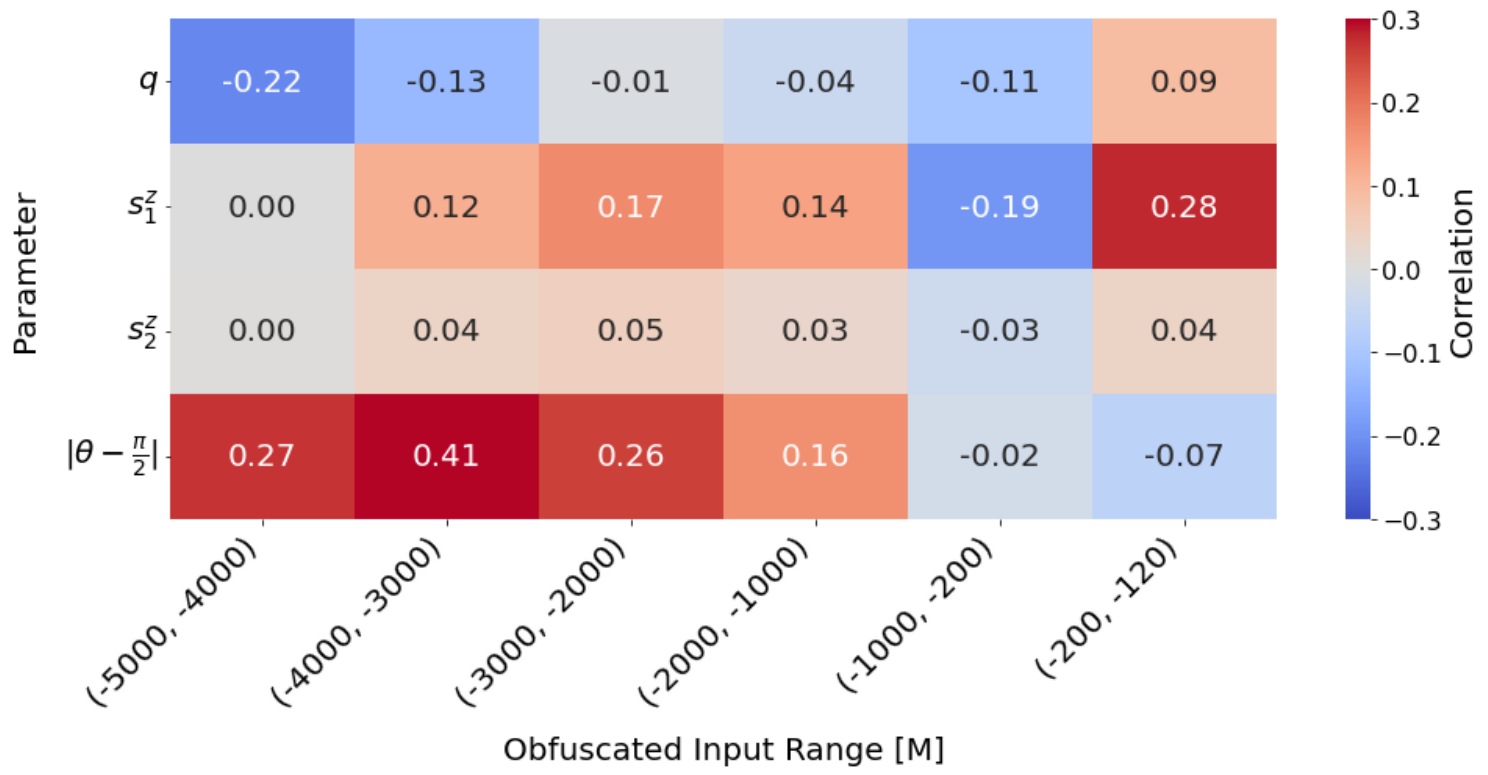}  
    \caption{Correlation matrix showing how waveform parameters correlate with the drop in overlap when different parts of the input are obfuscated. Each row corresponds to a physical parameter: mass ratio $q$, spin components $s_1^z$ and $s_2^z$, and the transformed inclination $|\theta - \frac{\pi}{2}|$, which captures the approximately symmetric structure observed around  edge-on orientations. Each column corresponds to a time window in units of $\mathrm{M}$.}
    \label{fig:correlation_matrix}
\end{figure}
    
    \item \textbf{Inclination Angle ($\theta$):} The inclination angle exhibits nontrivial correlation even in early inspiral segments, suggesting that the model leverages amplitude and polarization modulations induced by viewing geometry across long timescales. Such modulations arise because inclination affects how spherical harmonic modes project onto the line of sight: face-on systems are dominated by the $(2,2)$ mode, while edge-on systems excite a broader spectrum of harmonics. These geometric effects alter signal morphology across the entire waveform, as noted in \cite{PhysRevD.87.084008}. Notably, the match drop for edge-on binaries is near zero except in the final segments $(-1000\mathrm{M}, -120\mathrm{M})$ (see \ref{appendix:drop_bins}), likely due to the suppressed cross-polarization component at $\theta \approx \pi/2$, which reduces the effect of masking rather than indicating model disinterest. As such, low match drops near edge-on inclination should be interpreted with caution, as they may reflect vanishing waveform content rather than reduced importance.
\end{itemize}

Overall, these results offer quantitative insight into the localized importance of waveform features across time. Across all binary configurations, aligned-spin systems (i.e., systems where the primary black hole’s spin is aligned with the orbital angular momentum) and those with large $\theta \approx \pi$ are among the best resolved by the model and exhibit the highest information content in the late inspiral, as revealed by our obfuscation studies. In contrast, systems with anti-aligned primary spin or those with $\theta \approx 0$ carry less predictive information in the final input segments and correspond to regions of degraded performance. These patterns suggest a possible relationship between the informational richness of the late inspiral and the model's reconstruction accuracy: systems where specifically the late-time waveform evolution is more informative tend to be more successfully forecasted.

\subsubsection{Mode Modified Datasets}

To evaluate the contribution of higher-order gravitational wave modes to model accuracy, we constructed two additional test sets with truncated mode content. The first includes only modes up to $\ell_{\max}=3$, while the second is limited to $\ell_{\max}=2$. These reduced-mode datasets are generated using the same procedure and cover the same temporal domain as the full 
dataset, $t \in [-5000\textrm{M}, 130\textrm{M}]$, but exclude subdominant modes. Unlike in the obfuscation experiments, the ground truth here is adjusted to match the ablated input. This ensures that performance differences reflect the absence of subdominant modes rather than inconsistencies between input and target.

For each dataset, we compute the overlap score $\mathcal{O}$ between model predictions and the corresponding reduced-mode ground truth. Summary statistics are presented in Table~\ref{tab:overlap_modes}. As anticipated, the full-mode dataset, $\ell_{\max}=4$, yields the highest mean and median overlaps and the lowest standard deviation, whereas reducing mode content progressively decreases accuracy and increases variance. 

\begin{table}[ht]
\centering
\caption{Overlap statistics for different mode cutoffs.}
\begin{tabular}{lccc}
\hline
Dataset & Mean $\mathcal{O}$ & Median $\mathcal{O}$ & Std \\
\hline
$\ell_{\max} = 2$ & 0.9687 & 0.9890 & 0.0575 \\
$\ell_{\max} = 3$ & 0.9864 & 0.9938 & 0.0226 \\
$\ell_{\max} = 4$ (Full) & 0.9959 & 0.9974 & 0.0052 \\
\hline
\end{tabular}
\label{tab:overlap_modes}
\end{table}

\begin{figure}[h]
    \centering
    \includegraphics[width=0.99\textwidth]{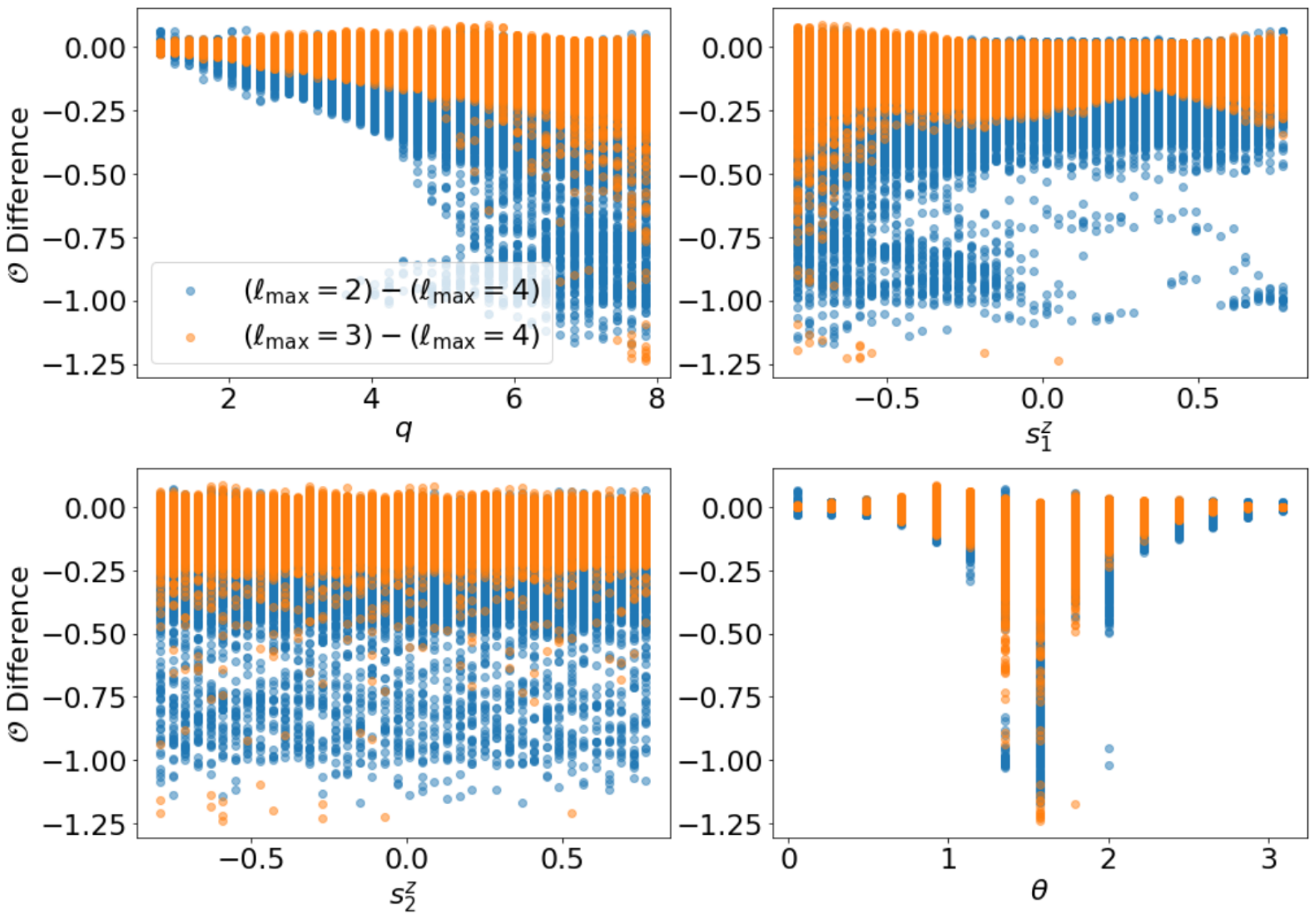}  
    \caption{Each point compares either ${\mathcal{O}}\left(\ell_{\max}=2\right)- {\mathcal{O}}\left(\ell_{\max}=4\right)$ or ${\mathcal{O}}\left(\ell_{\max}=3\right)- {\mathcal{O}}\left(\ell_{\max}=4\right)$ as a function of 
    binary parameters. The inclusion of higher order modes 
    improves forecasting accuracy for binaries with large mass ratios, low spin values of the heavier black hole, $s^z_1$, and edge-on orientations.}
    \label{fig:heatmap_modes}
\end{figure}

To understand where these discrepancies originate, we analyze the performance drop relative to the full-mode dataset as a function of intrinsic binary parameters. Figure~\ref{fig:heatmap_modes} displays the  loss in overlap for each reduced-mode variant across mass ratio $q$, the aligned spin of the heavier black hole $s^z_1$, and inclination angle $\theta$.

Our analysis reveals that higher-order modes improve accuracy in three specific regimes:

\begin{itemize}
    \item \textbf{High Mass Ratio ($q \gtrsim 4$):} The exclusion of higher modes leads to significant overlap loss, particularly when reducing to $\ell_{\max}=2$. This aligns with the increased contribution of subdominant harmonics in asymmetric mergers. As mass ratio deviates from 1, symmetry suppression of odd-$m$ and higher-$\ell$ modes weakens, and subdominant modes grow in significance. Mode truncation most affects high-$q$ systems, where the $(2,2)$ mode no longer dominates \cite{Berti:2007fi}.

    \item \textbf{Anti-Aligned Primary Spin ($s^z_1 < 0$):} Systems where the primary black hole’s spin is strongly anti-aligned with the orbital angular momentum show somewhat increased sensitivity to mode reduction, particularly in a small subset of cases with $s^z_1 \ll 0$ when excluding $\ell>3$. This trend is modest, and further analysis indicates it does not strongly correlate with spin–spin alignment or with effective spin. The increased mode sensitivity may instead reflect a compensatory role of subdominant harmonics in systems with retrograde alignment, which are known to exhibit distinct merger dynamics and reduced inspiral duration compared to aligned configurations \cite{2006PhRvD..74d1501C,PhysRevD.82.124008}.

    \item \textbf{Edge-On Orientations ($\theta \approx \pi/2$):} Orientation effects are especially pronounced in the transition from $\ell_{\max}=4$ to $\ell_{\max}=3$, but not between $\ell_{\max}=3$ and $\ell_{\max}=2$. This suggests that modes with $3 < \ell \leq 4$ 
    encode angular structure essential for modeling polarization and amplitude variations in edge-on configurations. For such orientations, the dominant $(2,2)$ mode contributes less power along the line of sight, while subdominant modes  project more strongly. As a result, neglecting higher-order modes leads to the largest mismatches in edge-on cases, where accurate waveform modeling requires their inclusion \cite{PhysRevD.87.084008}.

\end{itemize}

In contrast, the spin of the lighter black hole $s^z_2$ continues to play a secondary role across all mode configurations. The consistently weak dependence on $s^z_2$ suggests that the model captures its influence through more global, lower-order features.

Overall, these results confirm that including higher-order modes improves model fidelity, particularly in regimes where subdominant harmonics contribute significantly to the waveform structure. Systems where higher-order modes are essential—such as high mass-ratio, edge-on mergers—exhibit both high sensitivity to mode removal and reduced baseline performance, suggesting that dependence on subdominant mode content correlates with reconstruction difficulty, likely due to the increased structural complexity and reduced dominance of the quadrupole mode in these regimes. The model’s sensitivity to angular effects and binary asymmetries further highlights the value of full mode content for achieving robust and generalizable predictions.

\subsection{Attention Plots}
\begin{figure}[h]
    \centering
    \begin{subfigure}[b]{0.49\textwidth}
        \includegraphics[width=\textwidth, trim=0 80 0 0, clip]{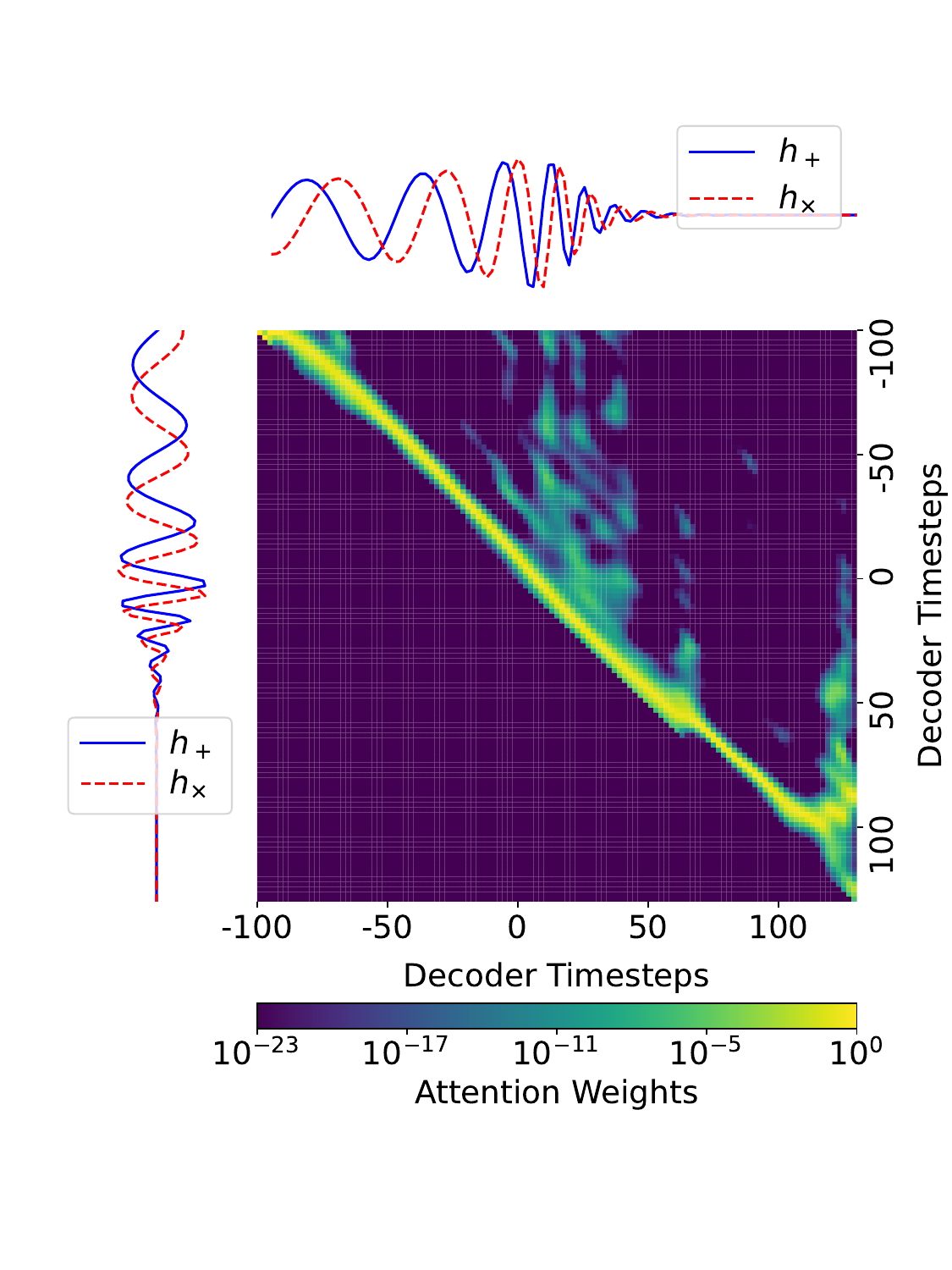}
        \label{fig:selfattention}
    \end{subfigure}
    \begin{subfigure}[b]{0.49\textwidth}
            \includegraphics[width=\textwidth, trim=0 80 0 0, clip]{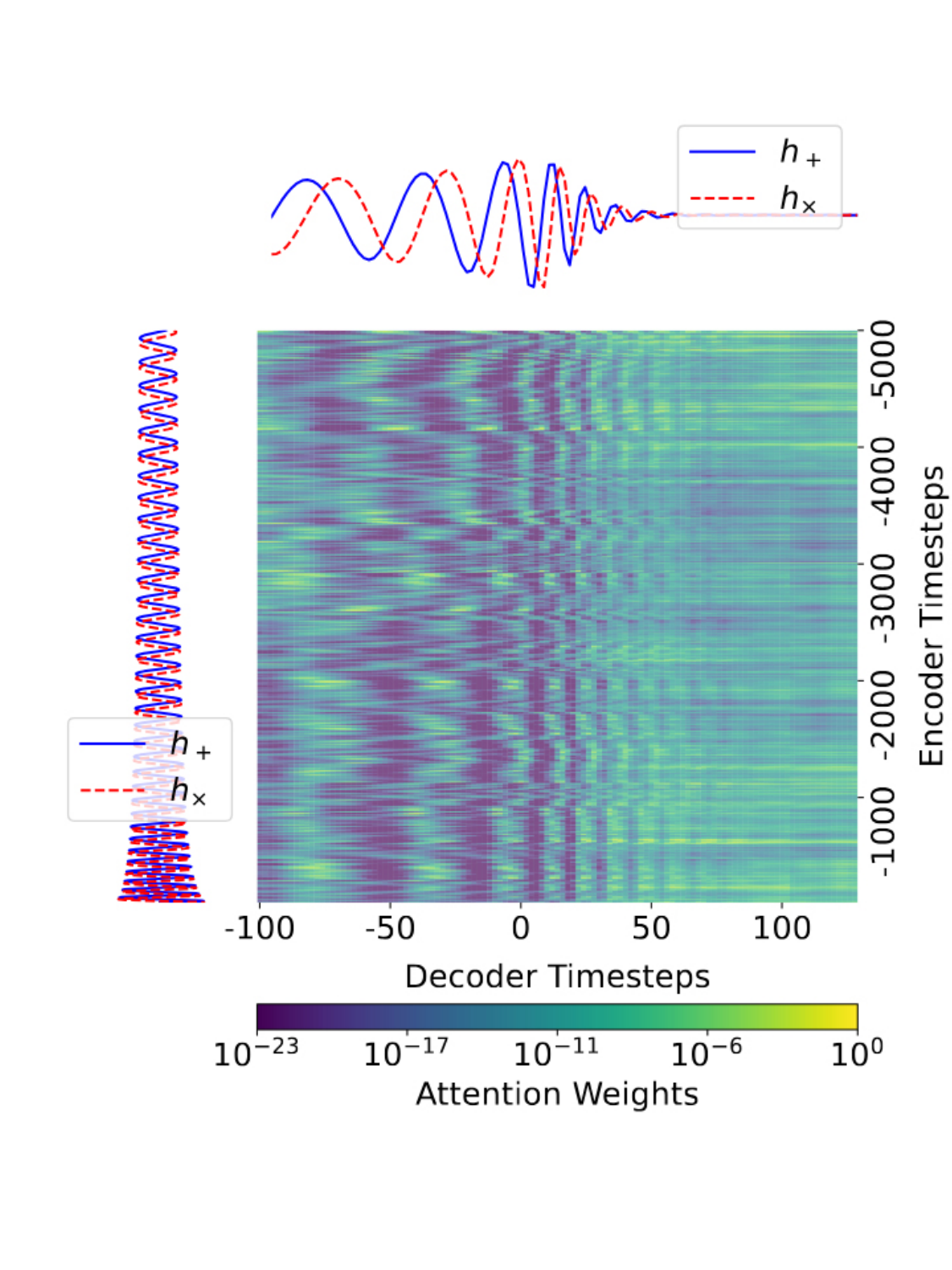}
        \label{fig:crossattention}
    \end{subfigure}
    \caption{Visualization of self-attention (left) and cross-attention (right) weights for a random choice of attention head and input waveform. The raw attention weights have been transposed to obtain the visual representation achieved here. To aid visualization, the corresponding input sequences have been plotted alongside the axes.}
    \label{fig:attention_comparison}
\end{figure}

In Figure~\ref{fig:attention_comparison}, we provide visualizations of the attention weights derived from the two decoder attention modules. While primarily qualitative, these visualizations help form an intuitive understanding of how the attention mechanism operates across different input waveforms and attention heads. 

As detailed in the model section, the attention weights are computed using the dot product of queries and keys, $Q K^T$, which measures the similarity between them. The softmax function then transforms these similarities into a probability distribution, resulting in a matrix of attention weights with shape $\mathbb{R}^{n_{\textrm{dec}} \times n_{\textrm{enc}}}$ in the case of the cross-attention module, or with shape $\mathbb{R}^{n_{\textrm{dec}} \times n_{\textrm{dec}}}$ in the case of the self-attention module. These weights are used to weigh the contribution of the values $V$ for the next representation. 

In both cross- and self-attention plots, we include the corresponding input sequences alongside the x- and y-axes to help visualize the model's focus. For self-attention, the plotted decoder input sequence represents both queries (x-axis) and keys (y-axis). For cross-attention, the plotted decoder input sequence corresponds to the queries (x-axis), while the encoder input sequence represents the keys (y-axis). Although the attention modules operate on embeddings or intermediate representations rather than the raw sequences shown, the plotted sequences provide a useful reference for interpreting the attention patterns. Note that each decoder timestep $t_{\textrm{dec}}$ in the cross-attention plot (i.e., each column in the transposed weight matrix corresponding to some decoder timestep $t_{\textrm{dec}}$) contains contextual information about how each encoder timestep relates to the current decoder timestep, which is needed for the transformer to make the $t_{\textrm{dec}}+1$ prediction in the decoder output. Due to the softmax operation, all values in each column sum up to 1.

The self-attention heatmap illustrates how each position in the decoder sequence attends to all other positions in the same sequence. For the self-attention mechanism in the decoder, each timestep can attend to all timesteps up to and including itself, but not to future timesteps, due to the causal attention mask. As a consequence of this property, there are no nonzero elements below the diagonal of the transposed weight matrix. As exemplified in Figure~\ref{fig:attention_comparison}, the diagonal elements frequently tend to dominate in these self-attention weight plots, indicating that local context is especially important to the self-attention mechanism when making predictions. The cross-attention heatmap reveals the model's focus on different parts of the encoder output sequence while decoding each position of the output sequence. The attention is more diffused than in the self-attention case, as the model is integrating information from a broader range of encoder timesteps to inform each prediction in the decoder output.

Interactive visualizations of the attention patterns for selected waveforms are available at ~\cite{githubhom} to support further qualitative inspection. However, we caution against overinterpreting these attention plots as a reflection of the model’s reasoning. The patterns tend to correlate closely with the amplitude structure of the input waveforms, showing little abstraction or parameter-dependent structure across examples. These findings are consistent with prior work indicating that attention weights do not necessarily provide meaningful explanations for model behavior\cite{jain2019attention}. Accordingly, we present these visualizations as diagnostic and instructive tools rather than interpretability claims.

\section{Discussion}
\label{sec:end}

This work demonstrates that transformer-based models can serve as fast, accurate surrogates for gravitational waveforms with higher-order modes, enabling waveform forecasting from quasi-circular, spinning, non-precessing binary black hole mergers across the late inspiral, merger, and ringdown phases. The model was trained on over 14 million waveforms using distributed training across 16 NVIDIA A100 GPUs on the Delta supercomputer, reaching convergence within 15 hours. Inference on 840,000 test waveforms was conducted using a single NVIDIA V100 GPU on the HAL system, with an average latency of 18 ms per waveform.

The model achieved high predictive accuracy across most of the signal manifold, with mean and median overlap scores of 0.996 and 0.997, respectively, and over 92$\%$ of samples out of the surrogate-based test set exceeding an overlap of 0.99. We also found that the model's conditional masking mechanism for the $h_{\times}$ polarization preserved accuracy in configurations where that signal component vanishes. 

To assess generalization beyond the surrogate training domain, we benchmarked the model against numerical relativity waveforms from the SXS catalog. Despite including systems outside the training domain and additional harmonic content, the model achieved a median overlap of 0.969, with low-overlap outliers scattered only around $\theta=\pi/2$, indicating strong performance even on high-fidelity, out-of-distribution data. The high data-to-parameter ratio (40,000:1) and strong NR generalization suggest the model avoids overfitting. 

Further analysis revealed that the model relies heavily on late-inspiral information in systems with high spin and mass asymmetry, while earlier inspiral features proved more important for near-equal-mass binaries. The model also appears to extract inclination angle-dependent structure well before merger. Mode truncation experiments confirmed that higher-order modes significantly enhance accuracy, particularly in systems with large mass ratios and edge-on orientations. 

While the model performs well across most of the parameter space, further improvement is needed for specific configurations, including low-mass-ratio, face-on binaries and high-mass-ratio, edge-on mergers—cases in which waveform complexity and amplitude suppression may challenge autoregressive modeling. These edge cases represent a small fraction of the test set, with only 0.003$\%$ of waveforms yielding overlaps below 0.90.

This work provides a starting point for future research into higher-order wave modes, including those from spinning, precessing binary black hole mergers. This broader signal manifold offers an opportunity to systematically explore the capabilities of AI models and exascale computing in learning complex, nonlinear dynamics of compact binary systems. These directions should be pursued in future work to further deepen our understanding of gravitational wave signals.

\section*{Acknowledgements}
\noindent This work was supported by Laboratory 
Directed Research and Development (LDRD) funding from 
Argonne National Laboratory, provided by the Director, 
Office of Science, of the U.S.\ Department of Energy under 
Contract No. DE-AC02-06CH11357, and by the Diaspora project 
of the U.S.\ Department of Energy, Office of Science, 
Advanced Scientific Computing Research, under 
contract number DE-AC02-06CH11357. 
KP was in part supported by the NSF grant NRT-1922512. 
The work used resources of the Argonne Leadership 
Computing Facility, a DOE Office of Science User 
Facility supported under Contract DE-AC02-06CH11357. 
EAH acknowledge support from 
National Science Foundation (NSF) award OAC-2209892. 
This research also used the Delta advanced computing and 
data resources, which is supported by the National Science 
Foundation (award OAC 2005572) and the State of Illinois. 
Delta is a joint effort of the University of Illinois 
Urbana-Champaign and its National Center for Supercomputing 
Applications. This research used the DeltaAI advanced 
computing and data resource, which is supported by the 
National Science Foundation (award OAC 2320345) and the 
State of Illinois. DeltaAI is a joint effort of the 
University of Illinois Urbana-Champaign and its 
National Center for Supercomputing Applications.


\pagebreak

\appendix
\section{Fine-Grained Analysis via Spin Slices}
\label{appendix:slices}

\begin{figure}[h!]
    \centering
    \includegraphics[width=0.9\linewidth]{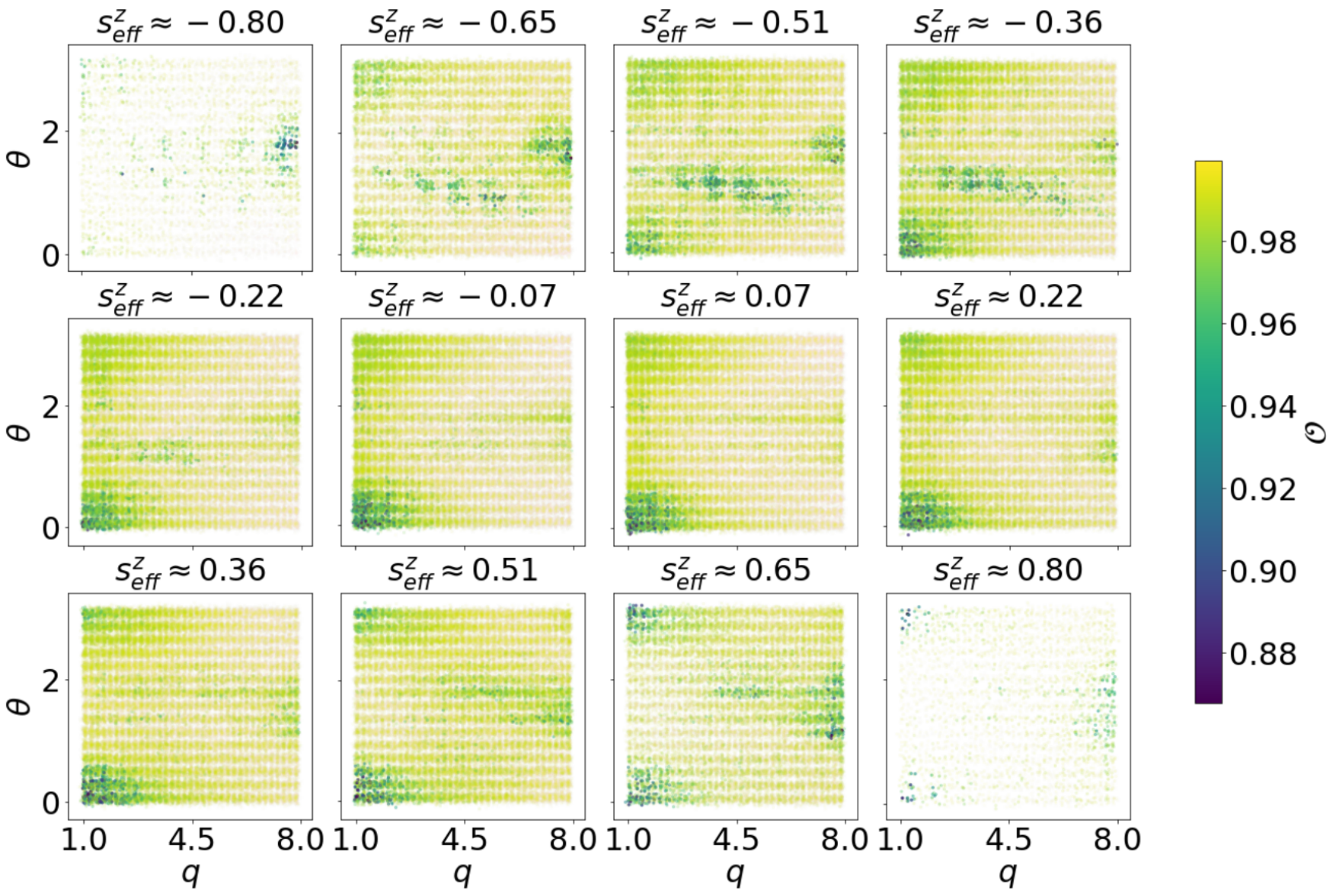}
    \caption{Reconstruction accuracy across mass ratio, $q$, and inclination angle, $\theta$, for several fixed values of the effective spin parameter, $s_{\mathrm{eff}}^z$.}
    \label{fig:spin_slices}
\end{figure}

Figure~\ref{fig:spin_slices} shows the test set partitioned into 
fixed-spin slices across the $(q, \theta)$ plane for various 
values of effective spin, $s_{\mathrm{eff}}^z$. This 
stratified analysis reveals that the transformer model 
consistently achieves high predictive accuracy for binaries 
with strongly aligned or anti-aligned spins, 
i.e., $s_{\mathrm{eff}}^z \sim \pm 0.8$. Within these regimes, 
the model performs robustly across a wide range of mass ratios 
and inclination angles.

However, at intermediate spin values—particularly around 
$s_{\mathrm{eff}}^z \sim 0$—a localized degradation 
in performance is observed. Specifically, reduced overlap 
values are concentrated around systems with comparable mass 
ratios ($q \sim 1$) and face-on orientations ($\theta \sim 0$), 
suggesting a sensitivity of the model to configurations with 
low intrinsic spin alignment and minimal inclination-induced 
modulation.

As $s_{\mathrm{eff}}^z$ departs from zero in either 
direction, predictive accuracy systematically improves across 
the $(q, \theta)$ slices, highlighting the model's effectiveness 
in capturing spin-modulated waveform features. Nonetheless, the 
model exhibits a consistent dip in accuracy for edge-on 
binaries ($\theta \sim \pi/2$), irrespective of spin, indicating 
a challenging regime associated with vanishing cross polarization 
and potential waveform degeneracies.

Finally, while the model attains near-optimal performance 
for $s_{\mathrm{eff}}^z \sim \pm 0.8$, we identify a narrow 
region of reduced accuracy at $s_{\mathrm{eff}}^z \sim -0.8$ 
for high-mass-ratio, edge-on systems ($q \sim 8$, $\theta \sim 
\pi/2$). This subtle degradation suggests a complex interplay 
between spin orientation, geometric projection effects, and 
mass asymmetry in these extreme regions of the parameter space.

\section{Binned Overlap Drop by Parameter for Obfuscation Study}
\label{appendix:drop_bins} 

To complement the correlation analysis in section \ref{obfuscation_experiments}, we provide binned heatmaps of the average overlap drop as a function of parameter value and obfuscated input range. These plots offer a finer-grained view of parameter influence on predictions, allowing us to assess possible nonlinear or symmetric effects that may be obscured by Pearson correlation.

\begin{figure}[h]
    \centering
    \includegraphics[width=0.9\linewidth]{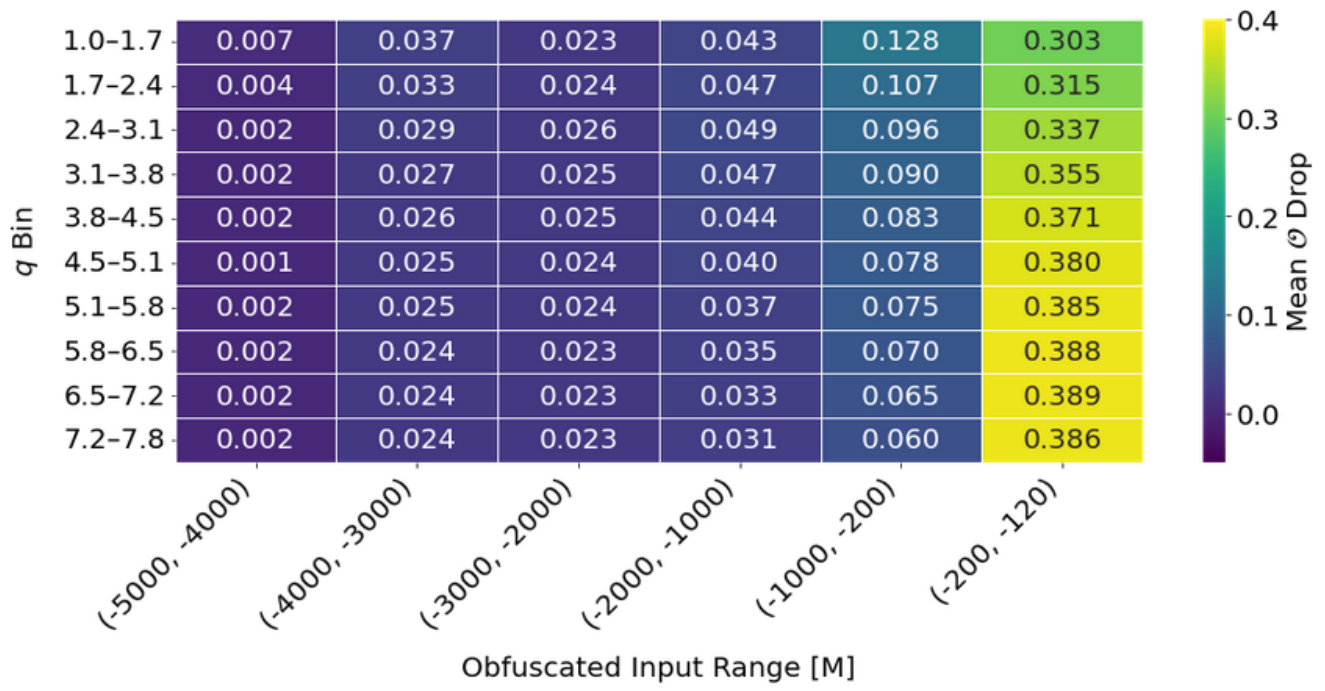}
    \caption{Binned overlap drop by mass ratio $q$. Most input segments show a broadly monotonic relationship, with match drop increasing for more asymmetric binaries (with the exception of the $(-1000\textrm{M},-200\textrm{M})$ range, where match drop decreases monotonically with $q$). Some early segments display small deviations from this monotonicity, particularly at low $q$. Notably, the low Pearson correlation observed in the earliest segment is largely driven by this localized effect around $q\approx1$.}
    \label{fig:qbin}
\end{figure}

\begin{figure}[h]
    \centering
    \includegraphics[width=0.9\linewidth]{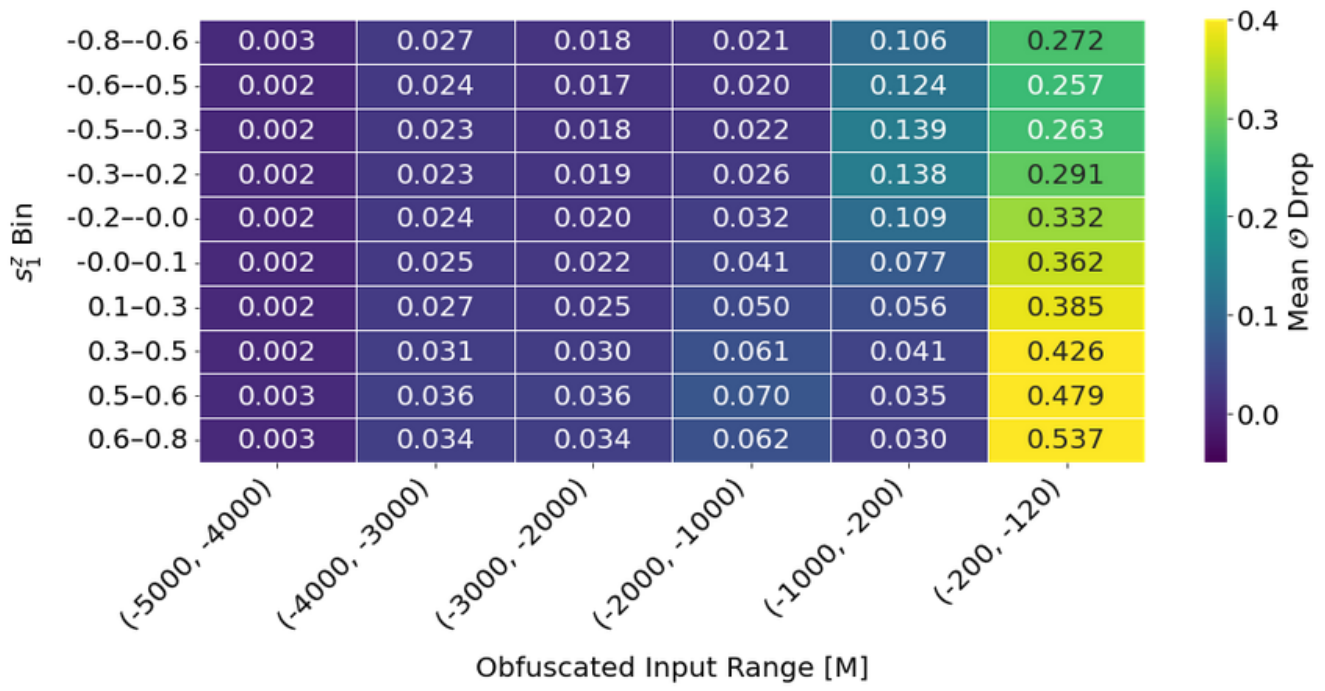}
    \caption{Binned overlap drop by primary spin $s_1^z$. The overall trend is approximately monotonic across the spin range, with match drop generally increasing with spin  (except in the $(-1000\textrm{M},-200\textrm{M})$ range, where match drop decreases with spin). A slight deviation from monotonicity is observed for very negative spin values $(s_1^z<-0.5)$, where match drop temporarily decreases before continuing the broader trend.}
    \label{fig:s1zbin}
\end{figure}

Figure~\ref{fig:qbin} shows match drop by mass ratio $q$. Most input segments exhibit a broadly monotonic increase in match drop with $q$, except in the $(-1000\textrm{M},-200\textrm{M})$ segment, where it decreases. Early segments show slight non-monotonicity, particularly near $q \approx 1$, explaining the low Pearson correlation in that region. Figure~\ref{fig:s1zbin} displays match drop by primary spin $s_1^z$. The trend is mostly monotonic, with drop increasing with spin. A deviation appears for $s_1^z < -0.5$, where drop temporarily decreases. In the $(-1000\textrm{M},-200\textrm{M})$ segment, match drop instead decreases with spin.

\pagebreak 

\begin{figure}[h]
    \centering
    \includegraphics[width=0.9\linewidth]{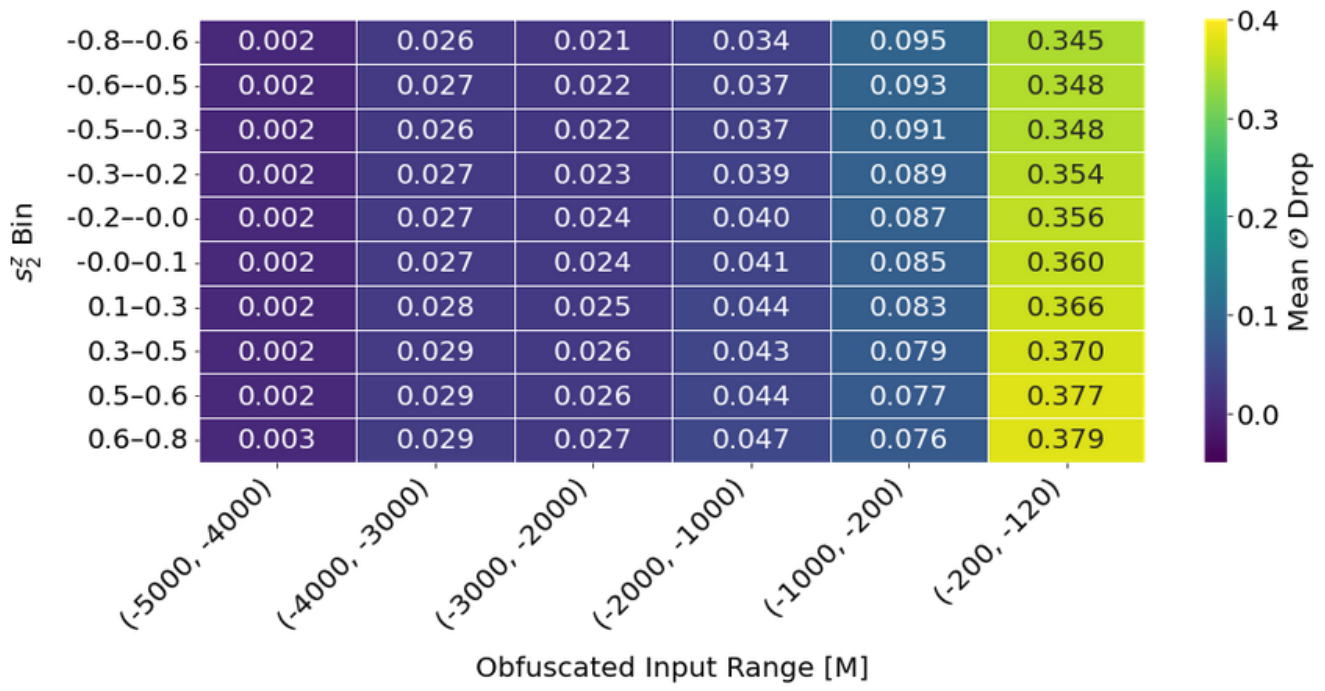}
    \caption{Binned overlap drop by secondary spin $s_2^z$. The overlap drop is relatively flat across all bins (i.e., within each column).}
    \label{fig:s2zbin}
\end{figure}

\begin{figure}[h]
    \centering
    \includegraphics[width=0.9\linewidth]{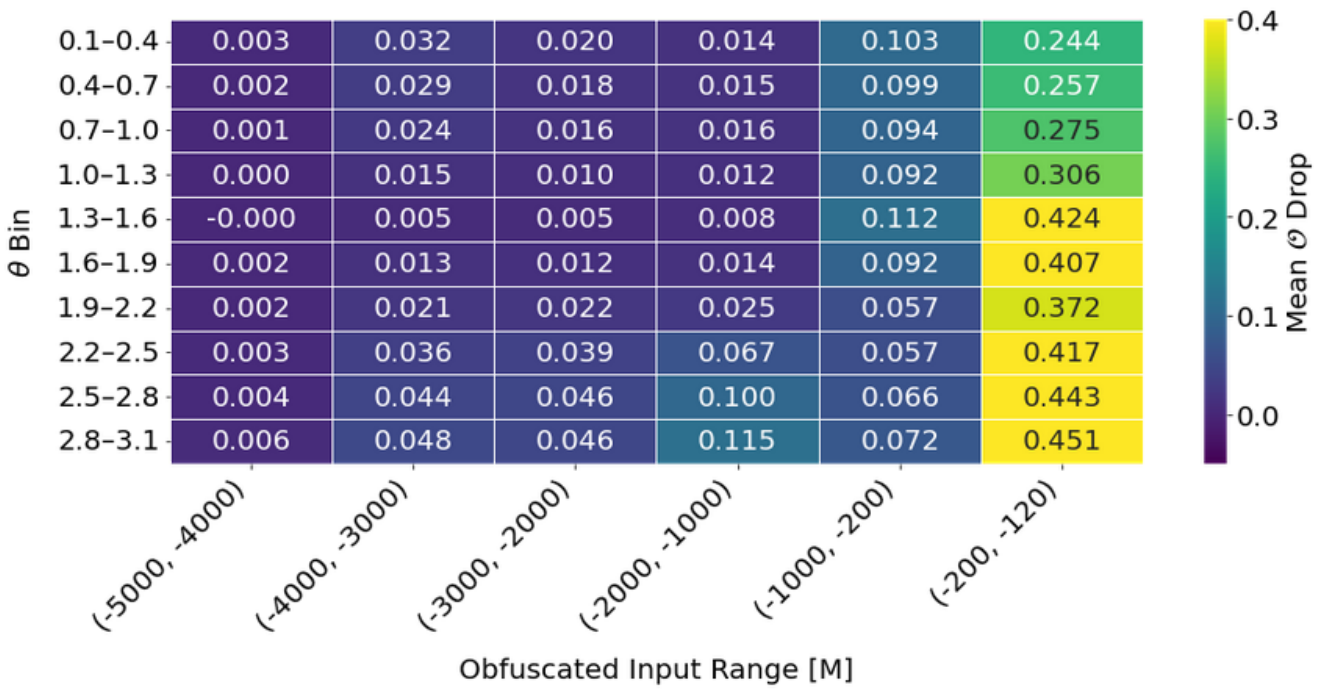}
    \caption{Binned overlap drop by inclination angle $\theta$. Most time segments show a roughly symmetric dependence around $\theta\approx\pi/2$, motivating the use of the transformed variable $\lvert\theta-\pi/2\rvert$ in the main analysis. In the final segment $(-200\textrm{M},-120\textrm{M})$, this symmetry breaks down slightly, with match drop increasing toward higher inclination angles.}
    \label{fig:thetabin}
\end{figure}

Figure~\ref{fig:s2zbin} shows match drop by secondary spin $s_2^z$, which is relatively flat across all bins.
Figure~\ref{fig:thetabin} plots match drop by inclination angle $\theta$. Most segments show a symmetric dependence around $\theta \approx \pi/2$, justifying the use of $|\theta - \pi/2|$ in the main analysis. In the final segment, this symmetry breaks slightly, with drop increasing toward higher inclination.

\pagebreak

\pagebreak

\section*{References}
\providecommand{\newblock}{}

\section*{Data availability statement} 
We produced training, validation and test datasets using 
the numerical relativity surrogate 
model \texttt{NRHybSur3dq8}~\cite{2019PhRvD99f4045V}, 
which is available at the following URL/DOI: 
\url{https://github.com/sxs-collaboration/gwsurrogate/tree/master}.  Additionally, we benchmarked the model's predictions against waveforms from the SXS catalog of full numerical relativity simulations, available at~\cite{SXSCatalogData_3.0.0}.

\section*{Code availability}
The scientific software to reproduce our results, including trained transformer models, and a tutorial for their use is available 
on GitHub~\cite{githubhom}.

\end{document}